\documentclass[10pt,journal,compsoc]{IEEEtran}
\ifCLASSOPTIONcompsoc
  \usepackage[nocompress]{cite}
\else
  \usepackage{cite}
\fi
\ifCLASSINFOpdf
\usepackage[pdftex]{graphicx}
\else
\fi
\usepackage{amsmath}
\usepackage{mdwmath}
\usepackage{mdwtab}
\usepackage{eqparbox}
\usepackage{algorithmic}

\usepackage{array}
\newcommand\MYhyperrefoptions{bookmarks=true,bookmarksnumbered=true,
pdfpagemode={UseOutlines},plainpages=false,pdfpagelabels=true,
colorlinks=true,linkcolor={black},citecolor={black},urlcolor={black},
pdftitle={Fairness and Privacy-Preserving in Federated Learning: A Survey},
pdfsubject={Typesetting},
pdfauthor={Michael D. Shell},
pdfkeywords={Federated learning; privacy-preserving; fairness; distributed machine learning}}
\ifCLASSINFOpdf
\usepackage[\MYhyperrefoptions,pdftex]{hyperref}
\else
\usepackage[\MYhyperrefoptions,breaklinks=true,dvips]{hyperref}
\usepackage{breakurl}
\fi

\begin{document}
%
\title{Fairness and Privacy-Preserving in Federated Learning: A Survey}
%
%
%
%

\author{Taki Hasan Rafi,
        Faiza Anan Noor,
        Tahmid Hussain,
        and Dong-Kyu Chae$^\dagger$
        
\IEEEcompsocitemizethanks{\IEEEcompsocthanksitem T. H. Rafi and D.K. Chae are with the Department of Computer Science, Hanyang University, South Korea.\protect\\
E-mail:\{takihr, dongkyu\}@hanyang.ac.kr
\IEEEcompsocthanksitem F. A. Noor was with the Department of Computer Science and Engineering, Ahsaullah University of Science and Technology, Bangladesh.\protect\\
E-mail: faizanoor.cse@aust.edu
\IEEEcompsocthanksitem T. Hussain is with the Department of Mathematical and Physical Sciences, East West University, Bangladesh.\protect\\
E-mail: tahmid.hussain@upaybd.com
\IEEEcompsocthanksitem $^\dagger$Corresponding Author.
\IEEEcompsocthanksitem Repo: \href{https://github.com/takihasan/Fairness-and-Privacy-in-FL-Survey}{\textcolor{blue}{https://github.com/takihasan/Fairness-and-Privacy-in-FL-Survey}}.
\protect\\
}}

\IEEEtitleabstractindextext{%
\begin{abstract}
Federated learning (FL) as a distributed machine learning strategy has gained immense popularity as privacy-aware Machine Learning (ML) systems have emerged as the solution to prevent privacy leakage by building a global model, and conducting individualized training of decentralized edge clients on their own private data. Prior existing works, however, employ privacy mechanisms such as Secure Multiparty Computing (SMC), Differential Privacy (DP), etc, which are immensely susceptible to interference, massive computational overhead, low accuracy, etc. With the increasingly broad deployment of FL systems (FLs), it is challenging to ensure fairness and maintain active client participation in FL systems. Very few works ensure reasonably satisfactory performances for the numerous diverse clients and fail to prevent potential bias against particular demographics in FL systems. The current efforts fail to strike a compromise between privacy, fairness, and model performance in FL systems and are vulnerable to a number of additional problems. In this paper, we provide a comprehensive survey stating the basic concepts of FL, the existing privacy challenges, techniques, and relevant works concerning privacy in FL. We also provide an extensive overview of the increasing fairness challenges, existing fairness notions, and the limited works that attempt both privacy and fairness in FL. By comprehensively describing the existing FL systems, we present the potential future directions pertaining to the challenges of privacy-preserving and fairness-aware FL systems.
\end{abstract}

}

\maketitle

\IEEEdisplaynontitleabstractindextext

%
\IEEEpeerreviewmaketitle

\ifCLASSOPTIONcompsoc
\IEEEraisesectionheading{\section{Introduction}}
\else
\section{Introduction}
\label{sec:introduction}
\fi

\IEEEPARstart{D}{ue} to the widespread adoption of personal gadgets, and edge devices in recent years, there has been an explosion in data generation, and a majority of this data are now spread across personal devices, different geographic locations, or institutions. Traditional machine-learning approaches rely on gathering raw data from multiple devices or institutions to a single central location (server or data center) in order to train models. Due to factors such as high communication costs, high computational power requirements, etc., collecting data at a central location is impractical in real-world scenarios \cite{huang2020fairness}. In addition, the security and privacy of end users or organizations are severely compromised when data from multiple sources are combined, sent to a centralized server, or shared directly between organizations or data centers \cite{liu2022distributed}. In such cases, a Collaborative Learning method, which does not involve collecting the users' private data, such as FL, can be effective and useful. FL is a distributed machine learning technique where multiple data owners, i.e., clients, collaborate together to accomplish learning tasks without the exchange of local data. This training scheme can utilize both centralized (server-controlled) and decentralized (server-less) settings. Since its inception in 2016, this learning strategy has been at the forefront of contemporary machine learning research. FL prohibits the transmission of distributed raw data between collaborating entities in order to ensure the confidentiality of user data. It only allows the participants to share intermediate data among themselves, which in most cases, is updated local gradients or models from the clients. Moreover, immediate aggregation of the updates (gradients or models) is performed as soon as feasible to provide additional protection against the disclosure of sensitive information \cite{10.1561/2200000083}. Besides, it also provides flexibility for the participants in the training process to join and leave the federation according to their own accord.

Nevertheless, recent research has revealed that FL systems may not always offer enough privacy and security guarantees for the participants. In general, security and privacy issues can be caused by a hostile server that intends to infer sensitive information from the local models' updates over time, meddle with the training process, or any adversarial participant that has the ability to deduce confidential data from other participants, manipulate the local parameter aggregation process, or corrupt the global model \cite{lyu2022privacy}. On the other hand, unfairness in FL systems may arise in different stages of the training process. Existing FL systems mostly suffer from unfairness issues from the following perspectives \cite{Shi_2023}:

(i) In the \emph{client selection} phase of FL, where methods prioritize the server's interests and exclude clients with inferior capabilities, unfairness may arise. This decreases the weaker participants' likelihood of obtaining a global model that corresponds well with their local data distributions. As a result, the global model may not generalize well, as clients ignored in the training process may contain data samples that are not covered by the present model.

(ii) \emph{FL incentive distribution} can also be unfair. Rewards may be needed in the FL training process to compensate for clients' resources and contributions. The FL system distributes the same global model to all clients as rewards under the standard FL configuration. However, such a strategy ignores the possibility that local client updates are frequently of varied quality due to differences in the quality of data samples and training capabilities. This scheme may be perceived as unjust by clients, who usually have a greater contribution than others to the global model's performance.

(iii) To ensure fairness in client selection and incentive distribution, \emph{contribution evaluation} is a crucial step in FL. Contribution evaluation may be based on self-reported information from the clients' side or observations of incremental model enhancement. Existing FL systems use client contribution as an important criterion for selecting high-quality clients, and incentive schemes frequently take client contribution evaluation results into consideration. Unfairness in the evaluation of client contributions can have far-reaching effects on the entire FL training process.

(iv) Last but not least, ensuring \emph{group fairness} in FL system has been a major concern in the past few years. Lately, it has been observed that the outcome of FL systems may be biased towards a population subgroup characterized by a sensitive attribute, such as gender, race, or ethnicity.

These privacy and fairness problems, if not addressed properly, might have a severe impact on the dependability of FL as an alternative to centralized ML systems. Unfair treatment and privacy threats may strongly discourage clients from participating in the FL training process. So, it is needless to say that ensuring privacy and fairness in FL systems is of paramount importance.

\textbf{Motivation.} As FL research has received a growing amount of attention in recent years, numerous survey papers on FL have been published. Various FL-related topics have been the focus of these surveys. Over the years, there have been a few attempts at surveys regarding growing privacy and fairness concerns in FL. However, there is a lack of enough high-quality surveys on these topics, particularly regarding issues of unfairness in FL. Till now, there is only one comprehensive survey that tried to provide a review of fairness concerns and measures in FL \cite{Shi_2023}. However, this survey did not provide a comprehensive analysis of the group fairness issue in FL, which has been a significant concern for FL systems in recent years. Some noteworthy surveys on security and privacy issues and privacy preservation approaches in FL have been published recently.  Although some privacy issues pertinent to FL were discussed in \cite{lyu2020threats}, there was no discussion on privacy-preserving strategies that could be implemented to guarantee privacy in FL systems. The surveys in \cite{10.1145/3460427}, \cite{lyu2022privacy} analyze in depth the privacy and security threats to FL systems, along with potential attacks and defense mechanisms. A few surveys, such as \cite{liu2022privacypreserving} have tried to provide an extensive review of Privacy-Preserving Aggregation approaches in FL, a key privacy-preserving technique adopted in FL systems.
Despite the fact that there have been distinct surveys on privacy and fairness issues, no survey till now has attempted to provide insights on both fairness and privacy concerns combined. This is our primary motivation behind this combined survey on privacy and fairness concerns in FL, as ensuring both fairness and privacy is of paramount significance for ensuring the quick adoption of Federated Learning systems in real-world scenarios.

\textbf{Our Contributions.} In particular, our contributions in comparison to earlier works are as follows:

\begin{itemize}
 \item To the best of our knowledge, our survey is the first paper that comprehensively reviews two dominant categories together, namely privacy-preserving and fairness in Federated Learning (FL).
\item We provide a broad outline of recent and existing privacy and fairness methods, challenges, and relevant works in the context of FL.
\item Our survey investigates privacy concerns in FL, evaluating the advantages and limitations of prominent methods used to ensure privacy in FL systems. Additionally, we summarize common metrics employed to evaluate privacy in FL systems.
\item We outline the concept of fairness in FL systems, identifying factors that contribute to bias and lack of fairness. Furthermore, we provide an overview of various aspects of fairness in FL, highlighting challenges, limitations, and evaluation metrics in depth.
\item Finally, we establish potential challenges and future directions for effectively preserving both privacy and fairness in FL.
\end{itemize}

\textbf{Organization of the paper.}
In Section 2, we provide an overview of the fundamental ideas behind federated learning (FL). Section 3 discusses the privacy challenges and adversaries that a typical FL system encounters, despite promising to be the ultimate tool for establishing privacy through decentralized training of ML systems. In Section 4, we explore the most common methodologies used to tackle privacy risks and challenges. Section 5 introduces the metrics for measuring the level of privacy established in an FL system. In Section 6, we introduce the concept of fairness in FL systems and present the plausible causes of bias creation among diverse clients or certain demographic groups in FL. Section 7 presents works that attempt to establish fairness in FL systems by eradicating the existing bias sources. This section also discusses different notions that aim to establish fairness from various perspectives. In Section 8, we describe how fairness in FL can be measured or quantified. Section 9 discusses the limited amount of work that equally emphasizes both fairness and privacy in FL. In Section 10, we highlight the challenges that current FL systems face regarding privacy and fairness, and we propose future research directions. An illustration of privacy and fairness in FL is shown in Figure 2.

\section{Federated Learning: An Overview}
In general, FL refers to a machine learning strategy that attempts to jointly train an ML model globally using model parameters across numerous clients. Its core methodology involves creating local models from local datasets and then using local clients to exchange parameters (such as model weights or gradients) to construct a global model. A typical FL system comprises two main parties. An illustration of an overview of FL is shown in Figure 1.

\begin{figure}[t] 
\centering {\includegraphics[scale=0.4]{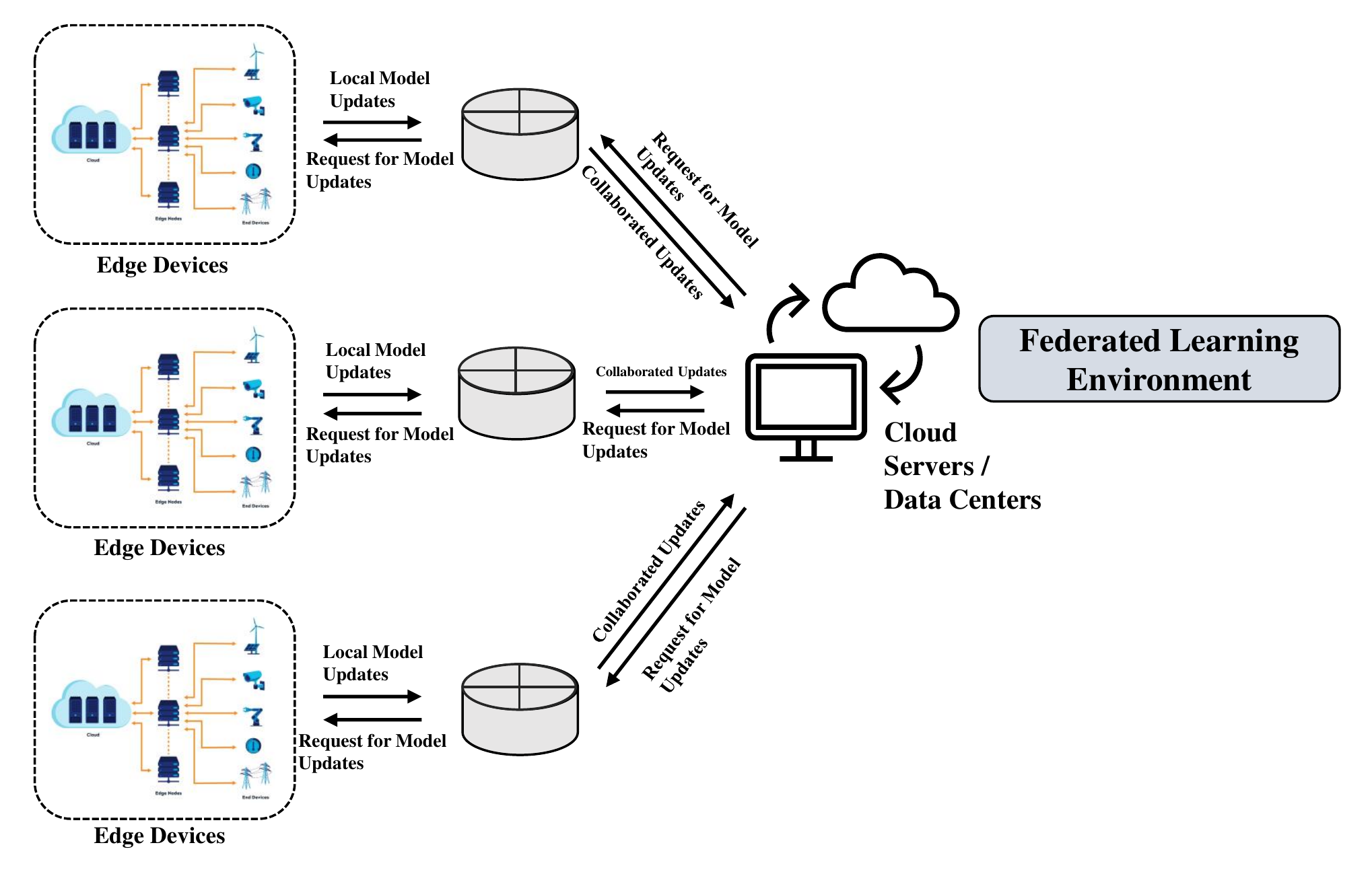}}
\caption{A typical overview of a federated learning environment.}
\end{figure}

\subsection{FL Categorization}
\textbf{Horizontal Federated Learning (HFL).}
  HFL is employed for situations where there is a small intersection on the sample space, but the various parties are involved to share the same feature space. Generally, most FL systems fall under this category and are prevalent in situations when each device possesses data containing identical feature space but varying sample space. Since all the local data exist in the same feature space, the clients have the advantage of using the same model architecture for training the models with that local data. By averaging all of the local individual models, the aggregated model can easily be updated using all the weights shared by the local models.\\
\textbf{Vertical Federated Learning (VFL).}
  Vertical Federated Learning, a less common category of FL, is employed in cases where each device has a dataset with unique characteristics derived from sample data. For instance, Vertical FL may be used to create a shared ML model across two organizations that have data on the same group of individuals but distinct feature sets. Since the local data lie in different feature spaces, it becomes difficult for the parties to share the same model architecture to train the local individual models with the same local data.\\
\textbf{Federated Transfer Learning (FTL).}
  FTL is appropriate when the feature space of the data varies together with the sample space. FTL normally exists as a mix of horizontal and vertical partitions. Most FL systems, however, commonly concentrate on horizontal partitioning. In case of instances where we only have enough training data for classification in one of our target domains, even when the classification task itself is in a different feature space or has a different data distribution from the training data, federated transfer learning \cite{5288526}, \cite{Liu_2020}, \cite{saha2021federated} can be used.\\

\subsection{FL Phases}
The four key phases of an FL system are \cite{liu2022distributed}, \cite{abdulrahman2020survey}, \cite{khan2021federated}: \\
\textbf{Client Selection Phase.} In this phase, the curator or global model or server either chooses the required clients at random from a group of devices or employs an algorithm for client selection. \\
\textbf{Broadcasting of Weights.}
After the server manages the mandatory task and data requirements and completes the hyperparameter specification, it broadcasts the global model weights or parameters. Additionally, it assigns work to the participants chosen in the client selection phase.\\
\textbf{Local Training.} This phase refers to the concurrent and individual training of each client, using each client's local data. \\
\textbf{Global Model Aggregation.} 
In this phase, each client, after completion of individual training, propagates their local model parameters to the server. Thereafter, the parameters are aggregated to construct the aggregated global model. These phases keep repeating until convergence.

\begin{figure*}[t] 
\centering {\includegraphics[scale=0.5]{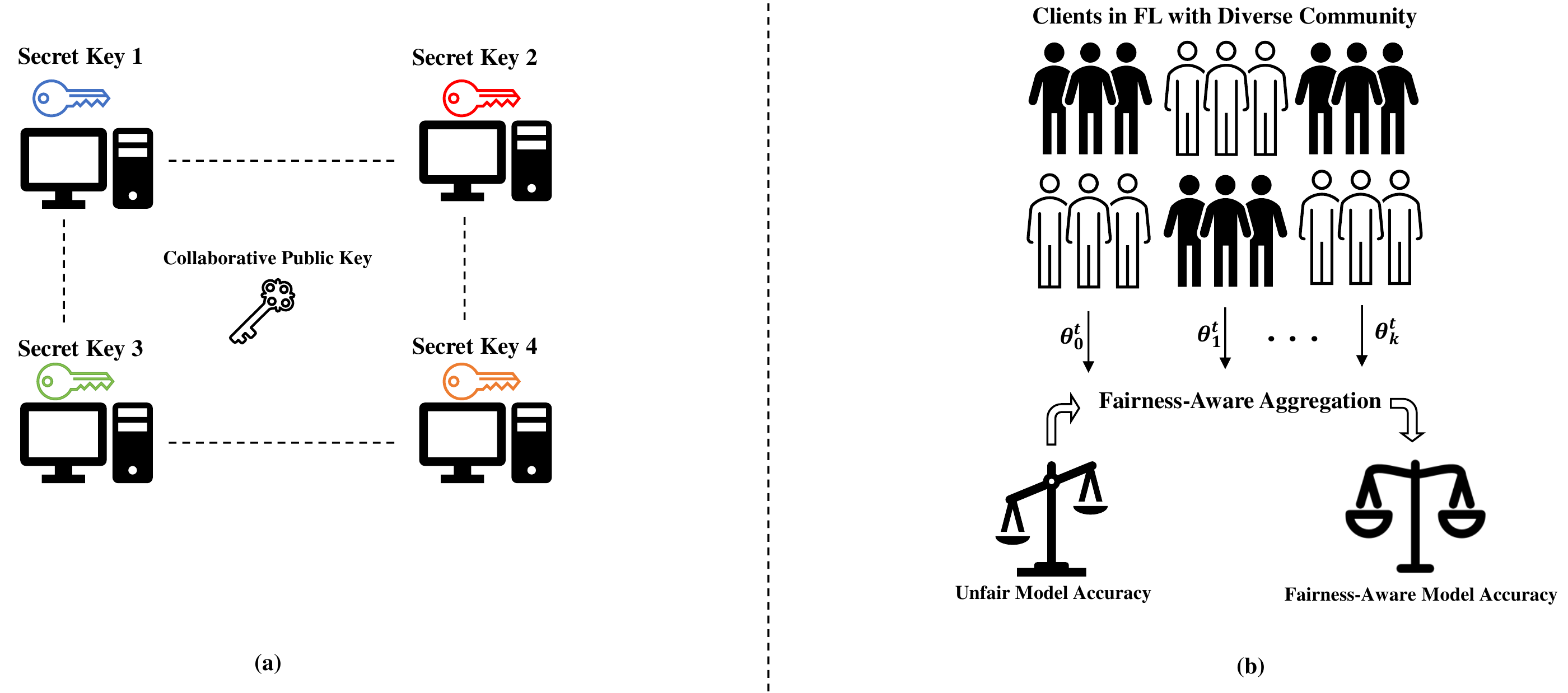}}
\caption{An Overview of Privacy-Preserving and Fairness in FL. (a) Privacy-Preserving in FL, and (b) Fairness in FL.}
\end{figure*}

\section{Privacy Challenges in FL}
The fundamental goal of a typical FL system is privacy preservation, as it enables systems or devices to jointly construct a global model while keeping all the training data locally on each individual system or device. Moreover, several clients perform concurrent local training on their own local data and communicate these gradients or weights procured during the local training phase to a central server. This process involves no private information in each client to be revealed, leading to a privacy-aware collaborative paradigm. Even though the primary goal is to hinder access to private or local data possessed by a particular client by other clients, several studies \cite{lyu2020threats}, \cite{mcmahan2018learning}, \cite{9109557},  \cite{wang2019beyond}, \cite{wu2022defense}, \cite{zhang2022privacy} demonstrate that FL systems are susceptible to several kinds of privacy attacks, primarily during the model weight exchange phase. A privacy attack often aims to deduce private data from training datasets, such as membership and class representatives and other properties. Model updates might leak extra information about unwanted properties of participants' private training data to hostile participants. Without possessing any prior knowledge of the training and regarding the model, an attacker can retrieve the original training samples by inferring labels from the shared gradients. Attacks made during the training in an FL model attempt to decipher, govern, or corrupt the entire model. The attacker might conduct model poisoning attacks or data poisoning attacks during the training phase to jeopardize the validity of the datasets used for training or to undermine the validity of the training procedure. The attacker is also able to execute a variety of inference attacks which are mainly evasion/exploratory assaults against a model's individual or collective updates. Attacks during the inference phase are attacks which aim on producing incorrect outputs and gathering information on the model's properties. Most privacy attacks occur during the inference phase. The inference attacks have the underlying assumption that the attackers possess unlimited computational resources, complex technical capabilities and full knowledge of the model. Inference attacks can be categorized mainly into the following types:\\
\subsection{Inferring Class Representatives}
Inference for class representatives attempt to produce examples that would not stand out in the primary dataset. The attacker can discover a lot about the data by effectively producing such samples. This kind of attack \cite{hitaj2017deep} leverages the advantage of real-time learning to enable the adversary to form a Generative Adversarial Network (GAN). This procures synthetic samples of the intended training set that was considered to be private, having the same distribution as the data reproduced.\\
\subsection{Inferring Membership}
Membership inference attempts to assess with accuracy whether a particular sample was used to train the network and acts on all target samples concurrently. An attacker, for example, can deduce if a given patient profile, such as a Pneumonia affected patient was used to develop a classifier that can distinguish between lung-affected patients. In the active membership inference scenario, the adversary has the ability to alter the FL model and launch a stronger assault on more participants. The target model is compelled to decrease the loss by descending in the direction of its local model's gradient by the gradient ascent attack. This can be used by the attackers to update the model parameters in the opposite direction of the loss gradient and can disseminate hostile updates and manipulate the model to reveal sensitive details about the local data of other players. In contrast, the attacker in the passive scenario observes the modified weights to make inferences and does not interfere with the learning process. \\
\subsection{Inferring Data Properties}
This sort of attack aims at deducing the meta-characteristics or the properties regarding the data being utilized. In case of an active adversary, multi-task learning can be leveraged and the FL model can be deceived to separate data with and without the target attribute in order to gather more data. Attacks of this type assume that supplemental training data are accurately labeled with the desired attribute by the adversary. A passive adversary can only keep track of the parameters and infer using a binary property classifier trained on them.\\
\subsection{Inferring Samples / Labels}
This type of assault or attack seeks to precisely replicate training samples utilized at the time of training and/or related labels that were used during training (rather than reconstructing them).  Deep Leakage from Gradients (DLG) \cite{zhu2019deep} follows an optimization approach, and shows how sharing the gradients might expose sensitive training data and reveals dataset properties such as the training samples and labels. This occurs with a minimal number of iterations. To carry out the attack,  a pair of synthetic or artificial inputs and labels are created. Then those dummy inputs and labels are optimized to diminish the gap between the artificial gradients and the original ones after deriving the artificial ones from the artificial data instead of maximizing model weights as in conventional training. GAN models \cite{wang2018inferring} can also be used to synthesize pictures that resemble training data from the gradient but the attack is constrained and only functions when the victim and adversary share at least one common label, which is impractical in real life. Privacy leaks and assaults may originate from one or more attackers \cite{8975792}. They may adopt the following forms:\\
\textbf{Single Adversary.} Privacy leakages can occur due to a single, non-colluding opponent \cite{lyu2020threats}, \cite{wu2020towards} such as a Federated client, the global server, or a third party even though a single attacker rarely possesses a central servers' attacking level.\\
\textbf{Colluding Attackers.} Even though attacks can occur with or without the help of the central server, collaborating FL users and main servers often increases the chance of privacy leaks and can cause more heinous attacks \cite{lyu2020threats}, \cite{panda2022sparsefed},  \cite{zhang2019pefl}.

The following two main categories can be used to categorize the enemies' capacity and role:\\
\textbf{Passive Malicious Server or Client (Honest-But-Curious or Semi-Honest).}
An FL global model or server can become curious and can retrieve clients' private data by observing the updates exchanged by the clients. Likewise, a curious FL client can retrieve private information about the system by observing the global parameters. These adversaries attempt to discover the personal information of honest individuals while adhering to the protocols of the system.\\
\textbf{Active Malicious Server or Client (Malicious while the Context is Clear).}
These adversaries or attackers alter identity or send false messages to other parties, but have a chance to deviate from the protocol of the system anytime. An actively malicious server is capable of performing strong assaults since it can also manage how each client views the overall model. The aggregated updates from all other clients are accessible to an adversary malicious client, who can construct their own malicious parameters. This allows a malicious FL server or client to obtain inner details about a client or a server's private data.

\section{Privacy-Preservation in FL}

The works on privacy preservation normally fall under four categories: 1) Cryptographic methods, 2) Anonymization methods, 3) Perturbative methods, and (4) Hybrid PPFL methods. Each of them are described below:
\subsection{Cryptographic Methods}
\subsubsection{Homomorphic Encryption (HE)}
By exchanging parameters inside the encryption method, HE safeguards user data privacy. In HE, data is encrypted into ciphertext using operations used to handle the original data.  The manipulation of the plaintext in this process is analogous to linear algebraic operations and does not involve decryption. The data or model are not exchanged, and can not be inferred from the data of the opposite party. For two messages, $M_{1}$ and $M_{2}$, one can compute En ($M_{1} + M_{2}$, $p_{k}$) using En ($M_{1}, p_{k}$) and En ($M_{2}, p_{k}$) without knowing anything about $M_{1}$ and $M_{2}$. Here, En ($\cdot$) refers to the encryption function and $p_{k}$ refers to the public key \cite{lyu2022privacy}. By using the secret key $s_{k}$ and the appropriate decryption function Dec($\cdot$), one can obtain $M_{1} + M_{2}$. This is an example of an additive Homomorphic Encryption mechanism that consists of the key pair ($p_{k}$, $s_{k}$). Currently, the existing methods of HE may be classified into the following categories: 1) Fully Homomorphic Encryption (FHE), and 2) Partially Homomorphic Encryption (PHE). FHE is comparatively more efficient and supports large ciphertext sizes, although it requires heavy computation resulting in a slow process. On the contrary, PHE schemes are less efficient and support small ciphertext sizes but is a swift process which requires less computational power. The most common techniques include  Paillier \cite{paillier1999public}, CKKS, and FV \cite{fan2012somewhat} schemes which are highly inefficient in terms of communicational and computational aspects. The CKKS technique adopts polynomial reductions and modular multiplications to encrypt real or complex numbers but only produces approximations.  The Paillier and FV algorithms both use integers as the plaintext. Even though the FV and CKKS methods offer both adds and multiplications on ciphertexts, the Paillier technique allows only additions to be done on encrypted data. An additively homomorphic encryption-based secure method \cite{kim2018logistic} is introduced for safeguarding the training and forecasting data using a logistic regression (LR) model and is implemented the system using encryption algorithms such as Paillier, LWE-based, and ring-LWE. They also underline the advantages and disadvantages of every implementation. The resulting system was incredibly scalable in terms of dataset size as well as dimension, supporting large sizes like hundreds of millions of records (108s). 

\subsubsection{SMC (Secure Multiparty Computation)} SMC \cite{bayatbabolghani2018secure} involves multiple parties or clients with private inputs and executes a collaborative computation on individual data or inputs without any sort of disclosure of private information. SecureML \cite{7958569} uses secure two-party computation (2PC) so that data owners can encrypt, process and secretly distribute their data across two servers that are non-colluding. Falling under the two server-model category, it enables every single client to train different models on their combined data without having to disclose any sort of data or information besides the results. However, this poses substantial computational and communication overhead, which may deter players will to collaborate. This method is a fault-tolerant, speedy, and safe protocol for secure aggregation suggested by \cite{bonawitz2016practical}. In this process, devices can exchange updates with the assumption that the service provider views it only after it has been combined by averaging it with the updates of other devices. For instance, an SMC protocol facilitates data sharing among servers (as $S_{1}, S_{2}, S_{3}$, etc.) in such a way that the data can only be reconstructed if the shared portions on n number of servers are known.  The advantage of training models using this technique is that it guarantees that the server only learns data about specific users and no party learns anything beyond the aggregate of the inputs from a small number of trustworthy users. SMC techniques frequently give a high level of accuracy and privacy at the expense of significant computing and communication costs, limiting their capacity to attract participants, and concurrent involvement of all parties becomes a major challenge. In exchange for efficiency, it is possible to build a security model with fewer security requirements using SMC. Partial knowledge disclosure may be considered permissible if security guarantees are provided. 

\subsubsection{Secret Sharing} Secret sharing \cite{beimel2011secret} refers to a cryptographic approach that ensures that a secret is made up of $N$ number of shares and can only be reconstructed if an adequate number of shares are joined. These techniques have been adopted by \cite{LIU202014}, \cite{8765347}, \cite{9005992}. t-out-of-n \cite{10.1145/359168.359176} technique demonstrates how to split a data (or secret) $S$ into $n$ parts so that $S$ may be readily reconstructed from any $m$ pieces, but even complete knowledge of $m - 1$ piece reveals no information about $S$. Using this methodology, strong key management schemes for cryptographic systems can be constructed that can continue to operate safely and reliably. SMC protocol was originally designed for securing aggregate gradients but its offline phase is quite complicated, and the system efficiency would be jeopardized if certain customers dropped out. Heterogeneous Federated Transfer Learning (HFTL) system \cite{9005992} uses a continuous privacy-preserving multi-party learning strategy with two variations based on HE and secret sharing mechanisms, respectively, and applied this technique for the case of mortality prediction in hospitals.  PFK-means \cite{LIU202014}, a privacy-preserving federated k-means technique attempts at providing proactive caching in cellular networks. This algorithm is built on FL and secret sharing. To cope with the backdoor attacks during training, a set of secret sharing protocols are utilized, and in the meantime, the data regarding client dropout is rebuilt using Lagrange basis polynomials to respond to the dynamic user change. This process is extremely time-consuming but is less computationally expensive and lighter than cryptographic methods such as existing HE or blockchain-based systems. A parameter server  \cite{shokri2015privacy} distributes the model gradients across the clients during training. Although they considerably increase efficiency, gradient leaks potentially compromise security. VerifyNet \cite{8765347} is an FL system that uses the secret sharing and key agreement protocol to safeguard the confidentiality of the participants' local gradients during the update phase.  To secure the privacy of users' local gradients during FL, they first used a double-masking approach. The cloud server then gives each user the evidence that the findings it has compiled are accurate. In this way, this scheme protects privacy and can be independently verified.

\subsection{Anonymization Methods}

Anonymization or de-identification mechanisms work by deleting any personally identifiable or detectable information from data. This could include a person's name, age, phone number, address, marital status, etc. These techniques have been employed by \cite{fung2019dancing}, \cite{9109557}, \cite{choudhury2020syntactic}, \cite{9325934}. TorMentor \cite{fung2019dancing} facilitates private multi-party ML, by expanding on FL. This system facilitates learning abstraction such that, in an untrusted environment, it enables data sources to contribute to a globally shared model with verifiable privacy guarantees. Multi-task GAN - Auxiliary Identification (mGAN-AI) \cite{9109557} combines GAN and a multi-task discriminator to concurrently distinguish between category and client identification of input. This system can retrieve user-specified private data and operates on the server side invisibly.
In addition, taking into account the anonymization approach for mGAN-AI mitigation, they suggests a linkability attack that re-identifies the anonymized updates by connecting the client representatives in advance. Furthermore, for determining the similarity of representatives, a brand-new Siamese network combining the identification and verification models is created.

Anonymization methods are classified into three types: k-anonymity \cite{sweeney2002k}, l-diversity \cite{mcmahan2018learning}, and t-closeness \cite{4221659}. These are specifically constructed for tabular datasets consisting of multiple records. QIDs are qualities of a dataset that, when combined, can re-identify persons through triangulation attacks. However, syntactic techniques for centralized settings have been much more thoroughly examined than the distributed FL scenario. The client side's original private data is anonymized using the k-anonymity algorithm as the first core component, and a global model is jointly trained using the anonymized data. A defendable degree of data privacy was provided by this approach such that it complies with regulatory requirements (such as the EU General Data Protection Regulation and the US Health Insurance Portability and Accountability Act). Unfortunately, k-anonymity cannot prevent linkage attacks in which a sensitive property is shared by a number of people who share the same QID. l-diversity extends k-anonymity to assure diversity among a group of people that share the same quasi-identifier \cite{bettini2006role}. t-closeness extends both of these strategies in order to retain the distribution of sensitive qualities across any group of individuals who share the same QID by diminishing the granularity of a data representation \cite{kiersztyn2021concept}. If the difference between the sensitive attribute distribution in an equivalent class and the attribute distribution over the whole table is less than a threshold $t$, the class is said to have t-closeness. A table is said to be t-close if all equivalence classes are t-close.  All techniques, however, suffer when an attacker has some knowledge of the sensitive property. Privacy leakage is reduced \cite{9325934} by allowing exchanging a lesser amount of parameters between the global model and clients. Differential privacy (DP) techniques on communicated parameters were applied with a Gaussian mechanism to achieve privacy. Moreover, to achieve participant anonymity, they used a proxy server as the intermediary layer between the global model and clients in order to lower the communication strain on the FL server. Ad-hoc anonymization might reasonably remove names, gender, phone numbers, social passwords, addresses, and other identifiers from each users' record(s), assuming that individuals cannot be identified within the changed dataset.

\subsection{Perturbative Methods}

Perturbation methods \cite{yang2021accuracylossless} are simple and effective methods that require no prior knowledge of the data distribution and it prevents clients from acquiring real global model parameters and local gradients. It works by incorporating arbitrary random integers as noises to the data, similar to DP. This causes the data derived from the disturbed parameters statistically to be identical to the original ones. PEFL \cite{9540342} uses long short-term memory-autoencoder approach and perturbation-based encoding to construct an FL-based gated recurrent unit neural network algorithm (FedGRU) for intrusion detection. These methods can be used to counter inquisitive clients' reconstruction and membership inference assaults by choosing random integers as perturbed noises added to the global model parameters, like Differential Privacy, rendering it extremely effective and simple to use practically.  But unlike Differential Privacy, Perturbation methods have no negative impact on learning accuracy as the server can retrieve the real gradients by deleting the introduced noises. However, these techniques are still susceptible to probabilistic privacy attacks. Mostly three kinds of separate perturbation techniques are often employed currently which are: 1) Differential privacy, 2) Additive Perturbation, and 3) Multiplicative Perturbation. 

\subsubsection{Differential Privacy (DP)} Differential privacy (DP) \cite{10.1145/3378679.3394533}, \cite{naseri2022local}, \cite{9714350}, \cite{hu2020personalized} is a technique that uses probabilistic statistical models to safeguard user privacy, and measures the level of private information disclosure of samples. Differential privacy mechanisms entail that privacy is achieved by introducing a small amount of Gaussian or exponentially distributed random noise or utilizing generalization techniques to hide specific sensitive qualities so that a third party is unable to differentiate the individual and the data can no longer be recovered. Even though the models are constructed after noise is injected, the DP approach is lossy, and can significantly lower performance in prediction accuracy, and the perfect tradeoff between accuracy and privacy is a challenge. DP can be categorized into the following categories:\\
\textbf{Centralized  Differential Privacy (CDP).}
This process was primarily meant for a centralized situation in which there exists a trusted or reliable server, that is authorized to access all clients' data and desires to respond to queries and publish statistics that protect user privacy by randomly generating query results. \\   
\textbf{Local Differential Privacy (LDP).}
When clients do not trust the server, LDP addresses CDP's flaws and assures privacy. In this process,  a differentially private alteration \cite{zhao2020local}, \cite{lyu2022privacy}, \cite{4690986} is applied to each client's data before sharing it with the server, and this allows for DP to be achieved in absence of the central server's participation. Because each party modifies its data independently, it is important to use LDP methods carefully to avoid having wrong predicted frequencies in the dataset. \\
\textbf{Distributed  Differential Privacy (DDP).}
In this process, the clients calculate and encrypt a brief, targeted report before sending the encoded reports to a secure computing function, the output of which is accessible to the central server. By the time the central server can view the output, it already complies with multiple privacy standards. To protect the clients' privacy, the encoding is carried out. By using safe aggregations and secure shuffling, this privacy-preserving method may be put into practice. \\
\textbf{Hybrid Differential Privacy (HDP).}
This process amalgamates various trust models by segmenting consumers according to their trust model choices. HDP-VFL \cite{wang2020hybrid}, the first ever hybrid differentially private (DP) framework for vertical federated learning (VFL), jointly trains a generalized linear model (GLM) from VFL with just minimal cost, compared to idealized non-private VFL. It relies on protocols like SMC and HE for privacy-preserving.

\subsubsection{Additive Perturbation Methods}
Additive perturbative techniques \cite{LIU2022103309}, \cite{9929413}, \cite{liu2008survey}, \cite{muralidhar1999general} are simple and quick perturbation techniques that require no prior knowledge of data distribution and aims to incorporate random noise into the data from a certain distribution (such as a uniform distribution or Gaussian distribution) in order to maintain the privacy of the original data while preserving the statistical properties. A class of methods \cite{chen2008survey} are discussed for privacy-preserving data mining. These methods randomly disturb the data using random value additive perturbation methods while maintaining its underlying probabilistic features. By introducing random noise, this method attempts to maintain the privacy of the data while ensuring that the random noise retains the signal from the data, allowing for accurate estimation of the patterns. Additive perturbation methods are less costly, easy to apply, and can be applied to each data point. Yet, it could reduce the value of the data and might be susceptible to noise reduction. 

\subsubsection{Multiplicative Perturbation Methods} A multiplicative perturbation \cite{yang2021learning}, \cite{1549830}, \cite{chen2008survey} method employs a set of transformation-invariant models, such as rotation and translation, so that they can be directly applied to the perturbed data, rather than introducing random noise to data. It converts the original data space into another space while keeping task and model particular information, obtaining the requisite model accuracy \cite{chang2019cronus}, \cite{Chamikara_2021}. Plain additive perturbation methods are not sufficient, a combination of additive and multiplicative procedures eradicate the drawbacks of both, as demonstrated by \cite{8619133}. POLAR-SGD (Private Optimization and Learning Algorithm with Stochastic Gradient Descent) \cite{Chamikara_2021} causes clients to hide the gradients via coupled additive and multiplicative perturbations. These obscured gradients are sent to several servers, which subsequently train prediction models. This research offers DISTPAB, a distributed perturbation technique for the privacy preservation of HFL, as a solution to these problems. By spreading out the burden of privacy using resource asymmetry in a distributed environment, DISTPAB reduces computational bottlenecks. Obscuring data at each IoT item using independent Gaussian random projection trains a Deep Neural Network (DNN) at the global model or server using the data projected from the IoT-based clients or objects \cite{LIU2022103309}. This method shifts the majority of the effort to the coordinator, who has access to adequate computing resources and introduces a minimal calculation overhead to the IoT objects. They conducted several studies using support vector machines and additive noise reduction for differential privacy but their comparative analysis demonstrates that this strategy outperforms the other methods.

\subsection{Hardware-Based Protection}
Hardware-level privacy assurances or trusted execution environments (TEE) such as secure processors embedded in client devices also exit, which can guarantee privacy in the event of operating system kernel security flaws. It is likely that such system-based or hardware-based privacy-preserving mechanisms will become more common owing to the growing influence of hardware-level ML-based works. Gradient preserving is obtained when using trustworthy SGX processors for this operation, even though server side channel attack exploitation still remains an issue. ShuffleFL \cite{10.1145/3457388.3458665}, a gradient-preserving system combines random group structure with intra-group gradient segment aggregation to defend against assaults using trustworthy SGX processors. Likewise, FLASH \cite{286415} also performs acceptable acceleration for cross-silo FL systems using hardware acceleration architecture. \cite{yang2020fpgabased} performs Pailier homomorphic Encryption using Hardware Accelerator for Efficient Federated Learning, possessing rigorous processing clock cycle, resource utilization, and clock frequency optimization for the modular multiplication operation. Flatee \cite{9581236}, a productive, privacy-preserving FL-based platform spans TEEs and significantly cuts down on training and communication time.

\subsection{Hybrid Privacy-Preserving Techniques}

Currently, hybrid privacy-preserving techniques approaches have been proposed to address the tradeoff between data privacy and data utility and to lessen computation and communication costs at the same time. The majority of FL systems employ the differential privacy approach which acts by perturbing values and ensures statistical indistinguishability for individual inputs.  Differential privacy adds a layer of randomization so that attackers with more knowledge still do not know the true value. Hence, this addition of randomization compromises accuracy and model efficiency while protecting user privacy. So differential privacy lessens the danger of data leakage but does not completely eliminate its occurrence and also jeopardizes model accuracy. \cite{mugunthan2019smpai}, \cite{10.1145/3029806.3029829}, \cite{choquettechoo2021capc} battled against data leakage and tried to upgrade model efficiency using the combined efforts of differential privacy and cryptography. SMPAI \cite{mugunthan2019smpai} uses SMC and DP and constructed models where the differentially private noise for each party is dispersed and created by other parties without any prior knowledge. \cite{choquettechoo2021capc} proposed Confidential and Private Collaborative (CaPC) learning and aimed to achieve both privacy and confidentiality in a collaborative situation, combining privately aggregated models with Secure Multi-party Computation (SMC), Homomorphic Encryption (HE), and other methods. Using this model, each party may increase model accuracy and fairness even in cases where each party has an overfitted model that produces good results on their own data only. This model is also suitable for situations where the data persists in a form that is not Independent and Identically Distributed(IID) and model architectures are variable across parties. This enables participants to cooperate without explicitly requiring them to connect their training sets or develop a central model. However, the computational costs are high.

\begin{table*}[!ht]
\centering
\caption{\textbf{A Summary on Privacy Techniques in FL.}}
\renewcommand{\arraystretch}{2}
\scalebox{0.75}{\begin{tabular} {p{3cm} p{5cm} p{4cm} p{4cm} p{5cm}}\hline 
\textbf{Privacy Preserving technique}&\textbf{Working Principle}&\textbf{Existing studies}&\textbf{Advantage}&\textbf{Disadvantage}\\    \hline 

HE & Cryptographic method that allows data to be processed mathematically as if it were unencrypted plain text, even though it has been encrypted.  & \cite{9812492}, \cite{ou2020homomorphic}, \cite{9952531}, \cite{ma2021privacypreserving}, \cite{zhang2020batchcrypt}  & The server is allowed to execute direct operations on encrypted data, and no decryption is required for operation executions.
 & The encryption operation is computationally expensive and time-consuming. \\  \hline

SMC & A group of methods and protocols that allow two or more parties to divide up data among themselves and do joint calculations without letting any one party see the data. &  \cite{mou2021verifiable}, \cite{li2020privacy}, \cite{xu2019hybridalpha}, \cite{liu2019enhancing}  & The parties involved have no restrictions on the control of the data owned. & Contains massive communication overhead. \\ \hline

DP & Obfuscation of individual data points in a dataset by modification or disruption while preserving limited interaction with the data. & \cite{wei2020federated}, \cite{triastcyn2019federated}, \cite{hu2020personalized}, \cite{seif2020wireless}, \cite{zhao2020local}, \cite{choudhury2019differential}, \cite{girgis2021shuffled}, \cite{adnan2022federated}, \cite{wu2022adaptive}, \cite{sun2020ldp}, \cite{kim2021federated}, \cite{rodriguez2020federated}, \cite{lian2021cofel}, \cite{wang2022safeguarding}, \cite{cao2020ifed}, \cite{basu2021benchmarking}, \cite{chen2022decentralized},
\cite{jia2021blockchain},
\cite{zhang2022understanding}, \cite{xin2020private}, \cite{li2019asynchronous} & Protection can be provided in the absence of prior background knowledge. & There are chances of unwanted noise that jeopardizes the quality and availability of the model.\\ \hline
Anonymization Methods & Elimination of information that is personally identifiable such as race, gender, or name.                                                                           &   \cite{choudhury2020anonymizing}, \cite{9770028}, \cite{chen2022federated}, \cite{10.1093/comjnl/bxab025}, \cite{angulo2022synthetic}.  & & Contains heavy computation and it is difficult to examine which quasi-identifier to use at each site. \\ \hline
Hardware-Based Techniques & In the case of operating system kernel security flaws, hardware-level privacy guarantees or trusted execution environments, such as secure processors can also provide privacy. & \cite{chamani2020mitigating}, \cite{10.1145/3457388.3458665}, \cite{9260194}, \cite{286415}\cite{yang2020fpgabased}, \cite{9908546}, \cite{chen2020training}, \cite{9581236} & This technique provides privacy and security in the case of operating system kernel flaws. & Highly expensive and difficult setup.\\ \hline
Hybrid Privacy-Preserving Techniques & This technique combines two or more privacy-preserving mechanisms. & \cite{truex2019hybrid}, \cite{jiang2021privacy}, \cite{xu2019hybridalpha}, \cite{zhang2023privacy}, \cite{mugunthan2019smpai}, \cite{10.1145/3029806.3029829}, \cite{choquettechoo2021capc}  & This technique tries to find a perfect balance between computational overhead, security, privacy leakage, etc. \\ \hline
\end{tabular}}
\end{table*}

\section{Privacy Preserving Metrics}

The metrics employed to assess the effectiveness of privacy-preserving techniques generally assess the loss of privacy of sensitive data and properties of the models or the dataset. Different methodologies measure privacy in different ways but a single metric might not qualify as the optimal tool for measurement. Aggregating multiple metrics can be useful if the aggregation preserves each metric's strengths while lowering its inaccuracies. Metrics that assess privacy only based on specific data features, make the assumption that all sorts of adversaries or inference attacks would only rely on these qualities and metrics not considering other types of adversaries implicitly imply a weak adversary. The factors for choosing metrics should be in accordance with the adversary \cite{Wagner_2018}, \cite{Wagner_2017} states around eighty privacy metrics and provides four categorizations on privacy metrics such as Adversary models, Data sources, Input, and Output. On the other hand, \cite{5958033} considers the attacker's accuracy, uncertainty, and correctness as viable metrics for measuring the loss of privacy in an FL system.
However, \cite{Wagner_2018} aggregates metrics using techniques such as adding sensitivity ratings, normalizing metrics, and extending measures to new settings. Moreover, to verify the system security and privacy, membership inference attacks \cite{nasr2019comprehensive} or gradient inversion attacks \cite{9679121}, \cite{huang2021evaluating} can be conducted and it can be verified by deducting the attacker's/adversary's success ratio. These assaults \cite{nasr2019comprehensive} can be carried out in one of two ways:  (1) A black-box attack \cite{9458510} is one in which the adversary has access to nothing except the final output model.  (2) White-box attack \cite{jiang2022comprehensive}: the adversary gets access to every model parameters that were traded during FL.

\section{Fairness in Federated Learning}
To address the privacy challenges and inference attacks raised by traditional machine learning, federated learning works efficiently since it allows for a new implementation of distributed and decentralized private training over a large number of clients. Unlike traditional centralized learning, which gathers all local data samples and develops the model on a central server, federated learning trains local models on local data samples while local clients communicate parameters to construct a global model. However, because of this decentralized nature of Federated Learning and since training data in federated learning is frequently geo-distributed among many organizations, translating solutions for fair training is difficult and the models commonly become biased or unfair towards some protected groups or clients. The performance of a model may differ dramatically between devices due to the heterogeneity of the data in federated networks. If a particular model A is fairer than another model B, then the test performance distribution of model A is more uniform across the network than that of model B \cite{li2020fair}. In other words, the standard deviation of model A is less than that of B. The existing works on fairness have mainly designed fair machine-learning models in federated learning contexts. Some of the most common approaches attempt to make each client produce fair judgments for people regardless of their protected features such as gender and race, while others try making optimal client selection choices and evaluating client contribution properly. The majority of these efforts have focused mostly on client fairness, which aims to design algorithms that result in models that perform similarly across diverse clients. 

\subsection{Causes of Bias in FL}
The following causes exist for bias generation in FL.\\
\textbf{Conventional Bias Sources.}
Generally, local training in an FL system is synonymous with training a model in a traditional ML system. The fundamental drivers of bias \cite{kamishima2012fairness}, \cite{kamishima2011fairness}, \cite{abay2020mitigating} in FL systems are also the same for ML, which are underestimation, prejudice, and negative legacy. Since FL involves several training sets, each participant will contribute its parameters (gradients or weights or biases) to the global model via shared model changes, the fairness of the final model is impacted due to the interactions among the clients and the server throughout the training phase. \\
\textbf{Sub-Sampling, Party Selection, and Dropouts.}
The aggregator queries participants in an FL process at each round of training to deliver their model changes, which the aggregator subsequently incorporates into the global federated model. Modern FL mechanisms do not evenly query all participants. The ability to engage in one round of FL training may be associated with sensitive traits, resulting in unintended bias. If a corporation utilizes cell phone data to train an FL model, network speed might influence whether or not a user's data is captured, which is connected to socioeconomic status. \\
\textbf{Data Heterogeneity.}
Data may not always be distributed equally across devices in federated environments and has the potential of generating severe complications. If the training data is not disseminated evenly throughout the clients, i.e. if the data survives in a non-i.i.d. form, the convergence behavior is severely jeopardized. Even when all parties are queried, another difficult but understudied element of FL is that each party's underlying data may differ. For instance, a branch near a women's training institution would provide data that was largely made up of women, which would be significantly different from the general composition of the bank's client information. \\
\textbf{Fusion Methodologies.} 
An aggregator uses a method prescribed by a fusion algorithm to combine local model updates into the global federated model. Depending on whether the aggregator achieves an equal or weighted average, the fusion techniques employed by the aggregator to merge the parties' model updates may introduce bias. FL methods may weigh heavier contributions for populations with more data i.e., heavy consumers of certain items, increasing the effects of over/under-representing specific groups in a dataset. \\
\textbf{Systems Heterogeneity.}
If models differ in terms of resource limits in federated contexts, unforeseen issues may occur. The battery power, network connectivity, hardware, and desire to participate in devices may vary. Furthermore, in FL, parties may be capable of dynamic participation, which allows them to drop out of an FL process and return later for a variety of reasons, such as connection constraints. As a result, the relative and overall data composition may be continually changing, affecting how the global model learns the bias.

\section{Recent Works in FL-Fairness}
In this section, we cover notable papers on fairness in the context of traditional Machine Learning, followed by a discussion of constructing fair models inside an FL framework. The several types of fairness in an FL system include:\\
\textbf{Performance Fairness.}
Performance fairness aims to achieve equal accuracy distribution across participants. Various notions of fairness such as Accuracy Parity are based on this concept of fairness.\\
\textbf{Collaboration Fairness.}
Collaboration fairness focuses on giving greater rewards/incentives to individuals with greater contributions. Notions of fairness such as Contribution Fairness aim to balance between Privacy-Preservation, Fairness, and Accuracy.\\
\textbf{Model Fairness.} Model fairness aims at the protection of some specific characteristics while maintaining the perfect balance between accuracy and privacy preservation using FL. It helps to achieve overall model fairness despite having some unrepresented demographic groups (based on race, gender, class, etc).

Apparently, FL has focused on fitting a single global model $m$, to all local data in the network. The global objective is to solve the following equation:

\begin{equation}
    \min_m \: G(L_1(m), \: L_2(m),\: L_3(m)...\: L_N(m)),
\end{equation}  
Here $L_n(m)$ stands for the individual or local objective for device n and $G(\cdot)$ refers to the global model objective that aggregates the local goals from each device. $G(\cdot)$ is commonly specified in FedAvg \cite{mcmahan2018learning} as the weighted average of all the individual or local client losses, i.e., $ \displaystyle\sum\limits_{n=1}^N  f_nL_n(m)$,

Here $f_n$ is a positive weight so that $\displaystyle\sum n \:f_n(w) = 1$. 

\subsection{Fairness Challenges in Horizontal and Vertical Federated Settings}
\textbf{Conflict Between Fair Model Training and FL.} Due to the conflicting inherent notions behind FL and fair model training, horizontal federated fair model training is difficult. For performing fair model training, getting access to all parties' data and all the samples is necessary in order to accurately assess a model's fairness. In contrast, Federated Learning requires that any access to any party's private data should be prohibited in order to protect data privacy. Likewise, in Vertical Federated Learning, features are distributed among clients. To properly assess the fair model training, performance over all the features is mandatory but this violates the main idea of VFL. \\
\textbf{Lack of Appropriate Fairness Measurement.}
No well-defined fairness measurement has been developed for Horizontal Federated Learning. Access to all the data is needed for other fairness measuring techniques in ML, which is not possible in horizontal FL. \\
\textbf{Problem with Local Fairness Estimation.}
In HFL, implementing fairness requirements locally on each client would result in worse fairness performance or null solutions due to erroneous model fairness measurements derived locally and the consequent conflicts between local fairness constraints. In VFL, model fairness cannot be measured for (passive) clients who are unaware of the protected property. Measuring model fairness for (active) customers that know the protected characteristic results in poor fairness and accuracy owing to a lack of features. \\
\textbf{Unfairness Towards Groups.} In VFL, the final model might be unfairly biased against groups possessing sensitive attributes since one or parts of the active clients may end up protected attributes.\\
\textbf{Imbalanced Processing Resources.} During the training process of a fair model in VFL, performance issues arise when enforcing every member to produce a single local update per communication round. There exists a lack of synchronization as different members may update in different time frames. The training process becomes increasingly difficult by the uneven distribution of processing resources among clients with varying attributes. 

\begin{table*}[!ht]
    \centering
     \caption{\textbf{Comparison Between Existing Works Focusing on Privacy in FL\.}}
    \renewcommand{\arraystretch}{1.85}

  \scalebox{0.9}{\begin{tabular} {p{1.9cm} p{3.0cm} p{2.8cm}  p{2.8cm} p{1.8cm}}
        \hline 
        
        \textbf{References}	& \textbf{Name of Model(s)} & 
        \textbf{Implemented Model(s)}  & \textbf{Data Partitioning} & \textbf{Privacy Mechanism}\\
        \hline

\cite{yang2019parallel} &Logistic Regression FL  & LM          & Vertical & CM \\ \hline

\cite{karimireddy2021scaffold} & SCAFFOLD  & LM, NN                          & Vertical & CM \\ \hline

\cite{cheng2021secureboost} & DT  & LM          & Vertical & CM \\ \hline

\cite{bhowmick2019protection} &Local DP FL & LM, NN         & Horizontal           &  DP \\ \hline

\cite{zhang2019pefl} &PEFL & DT, LM, NN         & Horizontal           &  CP \\ \hline

\cite{Hu_2022}  & OARF  & NN                    & Horizontal    & DP, CM         \\ \hline
\cite{jiang2021flashe} & FLASHE  & -  &  Horizontal & HE  \\  \hline

\cite{mcmahan2018learning} & FL-LSTM  &  NN     & Horizontal           &  DP \\ \hline

\cite{geyer2018differentially} & Client-Level DP FL  & NN  &  Horizontal & DP\\  \hline

\cite{wang2020federated} & NN  & LM          & Vertical & - \\ \hline

\cite{xie2021efficient}    & FedXGB  & DT               & Horizontal    & CM       \\ \hline
\cite{ribero2022federating}  &FedRecSys   & LM, NN      & Horizontal & CM \\ \hline
\cite{10004844}  &SimFL & DT                         & Horizontal & Hashing \\ \hline
\cite{truex2019hybrid} & Hybrid FL  &  LM, NN, DT    & Horizontal           &  CM, DP \\ \hline
\cite{Liu_2022}    & FedForest  & DT                   & Horizontal    & CM       \\ \hline
\cite{9448383} & -  & -  &  Horizontal & DP, HE \\  \hline

      \end{tabular}}
  
\end{table*}
\subsection{Notions of Fairness in Federated Learning}
The following notions currently exist to achieve fairness in Federated settings:\\
\textbf{Accuracy Parity.}
This notion focuses on increasing the fairness/uniformity of model performance across different clients while maintaining average performance overall.  Generally, a fair classifier trained on centralized data has a better performance trade-off than a model built in Federated settings based on a simple FEDAVG-based fair learning method. The basic trade-off between fairness and accuracy has been examined in a number of works and there are various ways to address this issue of finding the optimal balance between accuracy and privacy via Federated Learning algorithms. Data heterogeneity is the primary reason behind FL inconsistent performance. Training a customized or personalized model mostly yield better results than training a global model. Due to the diversity of data in an FL system, accuracy and model efficiency may be compromised. Personalization is a logical strategy used to increase accuracy given the varying nature of data in an FL system across varying clients. FEDFB \cite{zeng2022improving} uses personalization by modifying the FEDAVG protocol and develops a privately developed fair learning algorithm using decentralized data and mimics centralized model training. \cite{pmlr-v97-mohri19a} uses a new minimax optimization approach for federated settings called Agnostic Federated Learning(AFL). In AFL, the model is optimized for almost any kind of target distribution that is produced by a combination of varying data distributions among clients. The trained model does not overfit the data by favoring any particular client over others as it enhances the performance of the worst-performing or poorest device. 
q-Fair Federated Learning (q-FFL), \cite{li2020fair} presents a unique optimization goal that promotes an equitable and uniform accuracy between devices in federated networks. They provide a communication-efficient approach for resolving q-FFL called q-FedAvg that is appropriate for federated networks.  Re-weighting the objective by giving devices with poor performance higher weights would be a logical way to accomplish fairness as this would cause the distribution of accuracy in the network to move in the direction of greater homogeneity. The device performance heavily relies on the trained model. As the weights can not be guessed before training, therefore, dynamic re-weighting must be done. The foundation of this idea was laid by the idea of fair resource allocation in the context of wireless networks. For  positive local cost functions $L_{k}$, if parameter q is greater than 0, the q-Fair Federated Learning (q-FFL) optimization can be defined as:
\begin{equation}
    \min_{\theta}  L_q(\theta)=\displaystyle\sum\limits_{k=1}^N \frac{p_{k}}{q+1}F_k^{q+1}(\theta)
\end{equation}
$L_k^{q+1}$ represents $L_k$($\cdot$) to the power of (q+1), where q is a factor that controls the fairness level we want to enforce.  A higher value of q indicates more favor against devices with greater empirical losses locally, $L_k(\theta)$. As a result, more uniformity is imposed on accuracy and fairness. Setting $L_q(\theta)$ to a big enough q results in traditional minimax fairness  \cite{sharma2022federated} as the device with the highest loss dominates the objective. 
Since in practice,  each device may create data using a different distribution, hence to account for this variability, approaches that develop customized, device-specific models such as $v_k$ where k belongs to K, throughout the network are commonly considered.
A scalable federated multi-task learning system called Ditto \cite{li2021ditto}, where the key concept is local training with regularization, motivates individualized models to as close as possible to the ideal server or global model. Ditto improves absolute test accuracy by 5\% while decreasing variation among devices by an average of 10\% across all datasets as compared to TERM (on clean data). 

 Ditto's local goal or objective is $L_k(v_k)$. This mainly aims to build a model using just device m's data and included a regularization term that pushes individualized models to be as near as possible to the best server or global model.
For every device $m$, the optimization objective of Ditto becomes:

\begin{equation}
\begin{array}{cl}
\min _{v_m} & h_m\left(v_m ; \theta^*\right):LF_k\left(v_m\right)+\frac{\lambda}{2}\left\|v_m-\theta^*\right\|^2 \\

\end{array}
\end{equation}

In this case, the hyperparameter governs the balance between the global model and local models.
If the hyperparameter value is assigned to zero, Ditto is limited to training only the local models. In the contrary, when the hyperparameter value is assigned to a large value, it manages to recover the global model objective.

To investigate Ditto's fairness, it was compared to TERM (Tilted Empirical Risk Minimization) \cite{li2021tilted}, which is an updates version of the q-FFL \cite{li2020fair} objective for fair federated learning. TERM can be used for enforcing subgroup fairness, mitigating the impact of outliers, and dealing with class imbalance.  To provide flexible trade-offs between fairness and accuracy, TERM employs a parameter t. Positive values of t can help promote fairness (for example, by learning fair representations) and provide variance reduction for better generalization.\\
\textbf{Collaborative Fairness.}
Collaborative Fairness refers to a sort of distributed fairness notion that puts emphasis on the idea that a client's compensation should be in accordance with its contribution to an FL system. This notion, however, is not concerned with improving the accuracy of the model. A highly contributing participant, according to this notion of fairness, should be given greater rewards and they may be rewarded with a better-performing local model. 

For $N$ number of total participating clients, let $w_{i}$ be the model's weight after being trained at a client $i$. At the end of round $t$, the FL system's contribution from client $i$, if $G(w_t^i)$ is the gain from client $i$ becomes:
\begin{equation}
   P_{i}=\frac{G(w_t^i)}{\displaystyle\sum\limits_{i=1}^N G(w_t^j)} 
\end{equation}

Regardless of contributions, the majority of works in FL let all parties obtain the same model during and at the end of each communication cycle. But practically,  not all participants contribute equally for numerous reasons such as varying types and quantities of data that each member contains, and due to the differences in system and computational capacities of each participant.
As a result, although updates from some participants may improve the model, updates from others may potentially worsen the model's performance. The fact that everyone has access to the same global model at the conclusion of the partnership, irrespective of their contributions is unfair and in a worst-case scenario, even free-riders may join the system and unjustly make use of the global model. 
The absence of fairness could make robust parties exhibit a reduced willingness to participate.

Two difficulties and questions must be resolved in order to apply this notion, such as:
\begin{itemize}
    \item How to quantify an agent's contribution to an FL system?
    \item What kind of compensation should be given to the agents in order to accomplish fairness in proportion to their contribution?
\end{itemize}

The current ways of evaluating FL contributions may be classified into the following types:
1) Self-Addressed data, 2) Individualized Assessment, 3) Shapley Value (SV), 4) Utility game, 5) Empirical Approach, and 6) Leave-Out-One method.\\
\textbf{Self-Addressed Data.}
Here, client contributions are assessed based on the data provided by themselves. Information about their local datasets' quality, quantity, collecting expenses, computational and communication resources that they pledge to FL  are included in this information. These approaches make the assumption that customers are competent and reliable enough to accurately analyze their own circumstances and report the facts. However, this presumption is not applicable in practical scenarios. \cite{Shi_2023} requires data owners to disclose publicly verifiable parameters about individual local datasets to the FL server (e.g., data quality and quantity, data collecting cost, etc.). Afterward, the server assigns scores to the clients based on the provided data. \cite{9291636} and \cite{kang2019incentive} employ self-reported information to design an FL incentive system based on Contract Theory \cite{liu2022contract}, \cite{kang2019incentive} in which the server synthesizes contract items and distributes them to data owners. Each contracted item includes the incentives as well as information on the clients' local data. \\
\textbf{Individualized Assessment.} Individualized assessment methodologies prioritize individual performance over the overall FL model performance and it assesses a client's contribution based on its success in individual activities. A reputation system is commonly employed to maintain track of an FL participant's past participation. Client selection and incentive distribution methods might benefit from reputation procedures that represent the participant's dependability and participation. Individual assessment techniques frequently use two assumptions: 1)  FL server and the FL clients both can be equally trusted, and 2) a participant with a local model comparable to models from other participants (or with the global model) is regarded to contribute more. In actuality, these two assumptions may not hold true as it is established that the server and clients in FL can misbehave and can be greedy.  Furthermore, in non-IID situations, clients often possess data with diverse or varying class distributions. \\
\textbf{Shapley Value.} Shapley Value (SV) \cite{ghorbani2019data}, an extremely computationally inefficient method with complexity O(N!), is a traditional method in cooperative game theory for allocating total profits produced by a group of players. Data Shapley is used to quantify data evaluation and machine learning training and it may be possible to use it to estimate the size of an agent's contribution \cite{nagalapatti2021game}, \cite{fan2022fair}, \cite{zheng2022secure}, \cite{xi2021batfl}. A datums' Shapley value is determined by averaging its marginal performance over all remaining subsets of data. The contribution of the agent may then be determined by adding up the Shapley value of all of its data. The Shapley value of an agent for a predetermined learning task might vary depending on the model that is selected and consequently, data Shapley becomes an unreliable indicator of an agent's contribution to a given federated learning assignment. \cite{wang2019interpret}, \cite{9006179} uses an approach that uses Shapley Values to provide precise feature significance for host features and a unified important value for federated guest features, balancing the model interpretability and data privacy in vertical federated learning. Hence, due to perceived unfair treatment, Shapley value-based reward systems sometimes can cause the self-interested agents to quit the partnership. \\
\textbf{Utility Game.} Utility game-based Federated Learning approaches for evaluating contributions are closely connected to profit-sharing systems, which are sets of rules that translate the usefulness that players contribute into incentives. There are three popular profit-sharing programs:
\begin{enumerate}
    \item \textbf{Egalitarian:} Each unit of utility created by a team is distributed equally among the team members. 
    \item \textbf{Marginal Gain:} A participant's reward is equal to the utility the team acquired when they joined the team.
    \item \textbf{Marginal Loss:} A participant's payoff is equal to the utility they lost when they left the team. 
\end{enumerate}

Among the three, the marginal loss scheme is the most widely utilized strategy in FL. \cite{wang2019beyond} used the marginal loss technique to calculate the contributions of various stakeholders in Horizontal Federated Learning(HFL). The influence measurements are implemented using an approximation approach. To decrease communication and computation costs, \cite{Nishio_2019} used a marginal loss approach to quantify the client contribution within a single FL training phase. \\
\textbf{Empiritical Approach.} Shapley Value, a data-based contribution assessment method, has generated encouraging results for FL contribution evaluation. Its promising scalability is hindered by its high processing costs and imprecision due to estimation. FedCCEA, proposed by \cite{shyn2021fedccea}, learns the data quality of each customer by building an Accuracy Approximation Model (AAM) using sampling data size. This technique approximates the client's contribution reliably and efficiently by utilizing the sampled data size, and it allows clients to participate by specifying the desired sizes of local data to be utilized during training. FedCCEA is now confined to simple FL tasks since AAM is built on a fairly rudimentary neural network architecture, rendering it less suitable for practical applications. \cite{zhang2020hierarchically} suggested using publicly verifiable characteristics of agents, such as task-related data volume, data range, data collecting cost, etc., to quantify participating agent contributions. And in terms of fair incentives, proportional reward scheme was suggested, in which agents judged more valuable receive more model updates. \\
\textbf{Leave-Out-One Method.} The Leave-Out-One(LOO) \cite{9165379} approach quantifies the level of variance in a model's performance or predictions if that datum was not utilized during training. The duplicate of a datum is assigned a zero value using the LOO technique. If two agents have the same data, the additional data is considered to have no value. Additionally, the LOO approach, similar to Shapley values is model-dependent. FL incentive schemes frequently use this notion to drive reward distribution and various works exist to solve this issue. 

Collaborative Fair Federated Learning (CFFL) \cite{lyu2020collaborative} uses reputation to push participants to converge to distinct models in order to achieve fairness without jeopardizing the predictive performance. A reputation list for each participant is maintained by the server in the CFFL framework, and it is updated based on how well each member's gradients are uploaded throughout each communication cycle. Similarly, in order to accomplish collaborative fairness and adversarial robustness concurrently via a reputation system, Robust and Fair Federated Learning (RFFL) \cite{xu2021reputation} assessed each participant's contributions through their uploaded gradients (using vector similarity), and maintains a reputation for each participant and finds non-contributing or malicious players to be eliminated. Based on elements that may be publicly verified, such as data volume and quality, they identified agents' contributions. Later, they created the Hierarchical Federated Learning framework, or HFL, which supports the idea of fairness since it rewards agents in accordance with their previously agreed-upon contribution levels. \cite{10.1145/3375627.3375840} formulated the payoff-sharing Federated Learning Incentivizer (FLI) system. By jointly maximizing the collective utility and reducing the inequality among the servers in terms of the payout earned and the waiting time for getting compensation, the strategy \cite{zhao2023truthful}, \cite{tu2021incentive} dynamically allocates a given budget among the servers in a context-aware way. Fair and Differentially Private De-centralized Deep Learning framework enables collaborative fairness in collaborative learning (FDPDDL) as they used a unique two-stage reputation system, using digital tokens, local credibility, and differential privacy to assure justice and privacy. To mitigate privacy leakage, Differentially Private GAN (DPGAN) in the initialization stage and Differentially Private Stochastic Gradient Descent (DPSGD) in the update stage were used respectively. 

\subsection{Client Selection}

By giving underrepresented or unrepresented customers a better chance to participate, this idea of fairness seeks to reduce the bias in an FL model \cite{9814669}, \cite{9443523}, \cite{9916164}, \cite{9272649}, \cite{huang2020fairness}, \cite{qu2021contextaware},  
\cite{Shi_2023} achieved this by setting specific requirements such as long-term fairness constraints.
Unfairness may occur during the client selection step. Most of the existing works for client selection in FL hardly take the server's interest into account. This may include factors such as increased convergence speed, or improved model performance. These works act in favor of clients who can reply fast or who can enhance the performance of the system overall and customers with lower capacities could not be allowed to participate in the FL system. Consequently, the likelihood of two things is lessened namely,  a) the likelihood that weaker customers will receive a model that accurately represents their local data and (b) the likelihood that weaker customers will benefit from any associated incentives. From the global model's standpoint, the final model might not generalize effectively since the omitted clients might have samples uncovered by the current model.  FedMarl \cite{zhang2022multiagent}, an FL framework that depends on trained Multi-Agent Reinforcement Learning (MARL) agents to carry out effective client selection as it tackles control issues.

In the FedMarl \cite{zhang2022multiagent} procedure, three building blocks make up the central server: the Model storage block, which stores and updates the global DNN model, the MARL block, which runs the trained MARL agents and determines which clients to select, and the Statistics collection block, which collects data on client device statistics like processing latency and communication latency. The following steps constitute the workflow:

\begin{enumerate}
    \item \textbf{Step 1:} N client devices are selected from the client device pool for each training cycle.

    \item \textbf{Step 2:}  The global Deep Neural Network(DNN) model is then copied to all the devices specified in the Model storage block.

    \item \textbf{Step 3:}  After doing the probing training, the client devices submit their probing losses to the MARL block.

    \item \textbf{Step 4:}  Meanwhile, the client devices send the Statistics collection block their probing training latencies. 

    \item \textbf{Step 5:} After receiving all probing losses from clients, MARL agents combine these losses with historical data from the Statistics collecting block.

    \item \textbf{Step 6:} The client selection decisions are made in this step

    \item \textbf{Step 7:}  The chosen client devices then locally complete the remaining training and transmit the model parameters or weights to the Model storage block which applies the weight changes to the global DNN model. 
\end{enumerate}

FedCS \cite{Nishio_2019}, a client-selection problem, which can efficiently operate FL while a mobile edge computing(MEC) framework operator actively controls the resources of diverse clients. This system considers the client's computational resource restrictions, which they solved in a greedy fashion. FedCS, in particular, establishes a deadline for clients to perform operations such as downloading, updating, and uploading ML models. The operator then picks clients so that the global model may combine as many model updates as feasible in constrained time periods, making the whole training process more efficient and reducing the time necessary to train ML models. This process defines how many clients engage in the training process and when each client needs to complete the process.
\cite{cho2020client} presents POWER-OF-CHOICE, a client selection procedure that can flexibly balance between convergence speed and solution bias and quantified how the selection bias directly or indirectly influences the convergence speed of a Federated model. POWER-OF-CHOICE techniques converge up to three times quicker and provide 10\% greater test accuracy than random selection. They presented the first federated optimization convergence study for biased client selection techniques and demonstrated how biased client selection works leads to faster error convergence when it works in favor of clients with larger local loss. In order to enhance learning performance, customers should be selected wisely by inspecting the correlations between their data. They proposed a context-aware Neural Contextual Combinatorial Bandit method(NCCB), that respects the combinatorial restrictions given by federated learning while tactfully handling the link between the extracted attributes and rewards. Their selection technique is divided into two parts: context feature extraction and client selection. To retain the correlation link among clients and decrease the profile space, separate clusters are initially generated based on these feature vectors in the client selection section. Oort \cite{273723}, a typical multi-armed bandit problem, tackles the client selection issue without depending on contextual knowledge. It rather depends on local client training to estimate the local data quality, resulting in a protracted convergence time. Thus, Oort prioritizes clients who possess the ability of fast training and good quality data in order to facilitate model accuracy. CS-UCB-Q algorithm \cite{9142401}, based on UCB policy and virtual queue approach, tries to guarantee that each client can partake in training during a specified proportion of the communication rounds for dealing with cases of data present in a non-IID situation and imbalanced datasets. \cite{9155494} introduced FAVOR, which employs a reinforcement learning method for device selection in addition to bandit-based techniques.
\begin{table*}[!ht]
    \centering
     \caption{\textbf{Various Fairness Notions in FL.}}
    \renewcommand{\arraystretch}{2}

\scalebox{0.9}{\begin{tabular} {p{4cm} p{6cm} p{2.5cm} }
        \hline 
        \textbf{Fairness Notion} & 
         \textbf{Main Idea} &
         \textbf{References}	 \\
        \hline 

Group Fairness  & Reduces discrimination against a certain demographic(such as age, gender, societal class) & \cite{ezzeldin2022fairfed}, 
\cite{9378043}, \cite{zhang2022unified}, \cite{10.1145/3531146.3533081}, \cite{10.1145/3523227.3546771}, \cite{salazar2022fairfate}, \cite{papadaki2021federating}, \cite{pentyala2022privfairfl}, \cite{9861177}, \cite{juarez2023you} \\
        \hline 

Accuracy Parity & This notion tries to establish the correct balance between accuracy and fairness & \cite{wu2022motley}, \cite{horvath2022fjord}, \cite{9833842}, \cite{munir2021fedprune}, \cite{10020554}, \cite{li2020fair}, \cite{pmlr-v162-marfoq22a} \\ \hline 

Contribution Fairness & This distributed fairness notion focuses on the idea that the clients' payment must be in accordance with its contribution & \cite{lyu2020collaborative}, \cite{xu2021reputation}

\\ \hline

Client Selection & This idea focuses on mitigating bias in a model by creating a suitable client selection mechanism and tries to enhance the probability of participation of underrepresented clients. & \cite{9814669}, \cite{9443523}, \cite{9916164}, \cite{9272649}, \cite{huang2020fairness}, \cite{qu2021contextaware},  \cite{Shi_2023}\\ \hline

Good-Intent Fairness & This idea reduces the variation of model accuracy and efficiency among all the groups by optimizing the group that performs the poorest. & \cite{mohri2019agnostic}  \\ \hline

Regret 
Distribution Fairness & This idea of fairness seeks to reduce the disparity in FL clients' regret about having to wait to obtain incentive payout, taking into account the length of time the owner has been waiting to get the complete compensation.  & \cite{yu2020fairness} \\ \hline

 Expectation Fairness& The regret distribution fairness serves as a foundation for this idea of fairness. Since incentive benefits are dispersed progressively over time, it seeks to reduce the disparity between clients at various times. & \cite{yu2020fairness} \\ \hline

      \end{tabular}}
\end{table*}
\subsection{Group Fairness}
Group fairness, a key concept in ML, refers to the diminished bias in a trained models' performance against specific protected demographic groups, which are specified based on sensitive characteristics of the population (e.g., gender, race). Unfairness against certain age groups can be caused due to various reasons. Unfairness regarding treatment may arise in many real-world situations where there might be inadequate data related to one or more groups, such that the overall model performance is affected. For instance, when different hospitals want to develop a machine learning model for medical diagnosis for classification between Pneumonia affected and unaffected patients, but they only possess limited training data from a few groups because of their under-representation in a particular geographic location, the overall model performance may get affected and the model might become biased towards the group whose data is present predominantly. Let us assume that each data point is associated with a sensitive binary property U, for instance, race or gender. The fairness of a binary prediction model can be assessed by comparing its performance to the underlying groups defined by their sensitive characteristics. Not much work has been developed successfully to demonstrate Group Fairness even though the concept has existed for quite a while in conventional ML. FedOPT \cite{reddi2021adaptive} improves upon FedAvg and results in collaborative training of a highly performing aggregate model, but causes bias similar to traditional machine learning settings because it discriminates against a particular demographic group.

\subsection{Processing Algorithms for Eradicating Group Fairness}
Some preprocessing algorithms \cite{Chakraborty_2020}, \cite{friedler2019comparative}, \cite{zhang2021omnifair} aimed to achieve group fairness in centralized Machine Learning. These are also gaining popularity in Fairness-Aware FL implementations. These can be grouped into three categories:\\
\textbf{Pre-Processing Algorithms.} In this algorithm, prior to categorization, data is pre-processed to eliminate prejudice following techniques such as reweighing and optimized pre-processing. \cite{calmon2017optimized}, \cite{yang2020fairness}, \cite{biswas2021fair} improves model fairness through data correction during training leveraging this technique.  Equalized odds post-processing \cite{awasthi2020equalized}, \cite{mishler2021fairness} altered the dataset features beforehand to generate fair synthetic data. \\
\textbf{In-Processing Algorithms.} In this process, the dataset is separated into three parts here: training, validation, and testing.
After training, the trained system is optimized and validated on the validation set before being applied to the test set.
Some well-known works in this regard are Adversarial Debiasing \cite{zhang2018mitigating}, \cite{darlow2020latent}, \cite{hong2021federated} and Prejudice Remover \cite{kamishima2012fairness}.  \\
\textbf{Post-Processing Algorithms.} Here, some of the class labels are modified to lessen prejudice after classification by following methods such as Reject option classification \cite{herbei2006classification}, \cite{bartlett2008classification} and Equalized odds post-processing \cite{lohia2019bias}. \cite{small2023equalised}.  \cite{kim2019multiaccuracy}, \cite{putzel2022blackbox} improves the results of the system to provide more accurate predictions using post-processing.\\
\textbf{Processing Algorithms in FL.}
PrivFairFL \cite{pentyala2022privfairfl} uses SMC and DP in conjunction with FL, and used pre-processing and post-processing to mitigate the contradiction between fairness and privacy in FL and allowed cross-device FL training group-fair ML models under strict and official privacy protections. FairFed \cite{ezzeldin2022fairfed}, a methodology that allows for flexible usage of various local debiasing techniques among clients is server-side and independent of local debiasing.

\begin{table}[!ht]
    \centering
     \caption{\textbf{Existing Studies on Group Fairness in FL.}}
    \renewcommand{\arraystretch}{2.3}

\scalebox{0.9}{\begin{tabular} {p{1cm}p{2cm} p{1.5cm} p{2cm} }
        \hline 
        \textbf{Reference} & 
         \textbf{Model Name(s)} &
         \textbf{Privacy} & 
         \textbf{Processing}\\
        \hline 

\cite{pentyala2022privfairfl} & PrivFairFL & DP + SMC & Pre-processing, Post-Processing \\ \hline
\cite{ezzeldin2022fairfed} & FairFed & SMC & In-Processing \\ \hline
\cite{papadaki2021federating} & - & SMC & In-Processing \\ \hline
\cite{kanaparthy2022f3} & F3 & - & In-Processing \\ \hline
\cite{zhang2022unified} & FMDA-M & - & - \\ \hline
\cite{rodríguezgálvez2022enforcing} & FPFL & DP & In-Processing \\ \hline
\cite{yue2022gifairfl} & GIFAIR-FL & - & In-Processing \\ \hline
\cite{10.1145/3534585} & FCFL & - &  In-Processing \\ \hline
\cite{rodríguezgálvez2022enforcing} & FPFL & DP + SMC & In-Processing \\ \hline
\cite{abay2020mitigating} & - & DP & Pre-processing, In-Preprocessing \\ \hline
\cite{hong2021federated} & FADE & - & In-Processing \\ \hline
\end{tabular}}
\end{table}

\begin{table*}[!ht]
    \centering
     \caption{\textbf{Existing Studies on Fairness-Aware and Privacy-preserving FL.}}
    \renewcommand{\arraystretch}{2}

\scalebox{0.8}{\begin{tabular} {p{2cm} p{3cm} p{3cm} p{4cm} p{3cm} p{5cm}}
        \hline 
        
        \textbf{References}	 & \textbf{Model name(s)} &
        \textbf{Fairness notion(s)} & \textbf{Privacy Mechanism(s)} & \textbf{Data Partitioning(s)} & \textbf{Remarks} \\
        \hline 

\cite{Galli_2023} & FPFL  & Accuracy Parity & DP & Horizontal \\  \hline

\cite{padala2021federated} & -  & Accuracy Parity & Metric privacy (variant of DP) & Horizontal \\  \hline

\cite{pentyala2022privfairfl} & PrivFairFL  & Group Fairness & DP, SMC & Horizontal & The proposed method is not compatible with all concepts of fairness and is not suitable for application be applied in every sector\\  \hline

\cite{xia2021vertical} & Cascade Vertical Federated Learning (CVFL)  & Group Fairness & DP, SMC & Horizontal \\  \hline

\cite{9223632} & -   & Contribution Fairness & DP & Horizontal & Mechanism design (MD) and DP are used for an incentive design based private Federated Learning system \\ \hline
\cite{9378043} & FairFL  & Accuracy Parity & Secure aggregation protocol, Deep multi-agent reinforcement learning   & Horizontal \\  \hline

\cite{qi2022fairvfl} & FairVFL  & Accuracy Parity &  Contrastive adversarial learning, DP & Vertical \\  \hline

\cite{zeng2022improving} & FedFB,  & Group Fairness & DP & Horizontal & This work lacks the establishment of a correct three-way tradeoff between fairness, accuracy, privacy. \\ \hline

\cite{10.1145/3472456.3473513} & Dubhe  & Client Selection  &  HE & Horizontal & - \\  \hline

\cite{wang2023fedeba} & FedEBA+  & Accuracy Parity  &  - & Horizontal & - \\  \hline

\cite{huang2022aflpc} & AFLPC  & -  &  Adaptive DP & Horizontal & - \\  \hline

\cite{passerat2020blockchain} & -   & Blockchain  &  - & Horizontal &  This mechanism involves discarding of data as a part of model poisoning attack prevention which yields loss of useful data. It also lacks proper reputation-based rewards system for ensuring fairness among clients. \\  \hline

\cite{9686048} & -  & Blockchain  &  - & Horizontal & - \\  \hline
\cite{ezzeldin2022fairfed} & -  & Group Fairness  &  SMC & Horizontal & - \\  \hline
\cite{li2021ditto} & -  & Accuracy Parity  &  - & Horizontal & - \\  \hline
\cite{papadaki2021federating} & -  & Group Fairness  &  SMC & Horizontal & This technique requires exploration of privacy attacks and robustness \\ 
\hline

      \end{tabular}}
    
\end{table*}

The majority of these techniques rely on a single, publicly accessible training dataset, which violates data privacy and infringes on one of the main goals of an FL framework. 

\subsection{Notions of Group Fairness}
At present, various notions of group fairness exist, such as Equal Opportunity, Statistical Parity. Let, the quantity of true positives be ($TP_{ad}$ and $TP_{dis}$) for the advantaged group (ad) and disadvantaged group (dis) respectively. And ($TPR_{ad}$ and $TPR_{dis}$) be the true positive rates and false positive rate ($FPR_{ad}$ and $FPR_{dis}$) for the advantaged and disadvantaged group respectively. \\
\textbf{Equal Opportunity.}
According to \cite{hardt2016equality}, Equal Opportunity states that the predictor is deemed fair if the real positive rate is independent of the sensitive characteristic U. In other words, the False Negative Rate (FNR) of the system is not affected by the presence of a sensitive attribute U, ie. $FNR_{U}$ =FNR, $\forall{u}$. Equality of opportunity tries to produce an equal probability that the probability of prediction using sensitive attributes is the same as non-sensitive attributes. Equal Opportunity difference (EOD) can be the difference between True Positive Rate (TPR) for each group. EOD can be stated in the following way:
\begin{equation}
     EOD= |TPR_{dis}-TPR_{ad}|
\end{equation}
\textbf{Statistical Parity.}
Statistical parity \cite{dwork2012fairness} awards the classifier for correctly categorizing each group(groups that contain sensitive property and groups that lack any sensitive property) as positive at the same rate.
Hence, a model is considered fair from the standpoint of statistical parity, if it correctly identifies positive cases for both samples with sensitive and normal attributes.
Statistical parity difference (SPD) can be defined as the variation or difference in the proportion of successful outcomes for each group. SPD can be defined as:

\begin{equation}
     SPD= |(\frac{TP_{dis}}{N_{dis}})-(\frac{TP_{ad}}{N_{ad}})|
\end{equation}

Fairness is achieved in the case of values that are closer to zero for the EOD and SPD. If EOD and SPD yield positive results, it indicates that the disadvantaged group outperforms the advantaged group.

\section{Fairness Evaluation Metrics}
Establishing a set of performance assessment measures is crucial for the long-term viability of fairness-aware federated learning systems in order that the benefits of different proposed ways can be evaluated relatively.
Different metrics are generally used by different fairness-aware federated systems and are suitable for different scenarios and applications.\\
\textbf{General Evaluation Metrics.}
Normally, accuracy is used as a general metric to quantify how well a model performs \cite{9247530}, \cite{lyu2020collaborative}, \cite{8995775}, \cite{hu2023federated} and how each client contributes to the whole model so that an appropriate client selection scheme can be built \cite{lyu2020collaborative}, \cite{hu2023federated}, \cite{shyn2021fedccea}.
In contrast to accuracy, efficiency measures the performance of training, either by the number of rounds required during the training process or by the total training time taken. Some works attempt to lessen the training time \cite{Nishio_2019}, \cite{li2020fair},  while others try to lessen the number of training rounds \cite{huang2022stochastic}, \cite{wang2021federated}, \cite{Huang_2020}.\\
\textbf{Evaluation of fairness.}
Fairness evaluation metrics such as Average Variance(AV), Pearson's Correlation Coefficient, and various distance metrics are also used to evaluate fairness in Federated systems. Average Variance (AV) typically demonstrates how a particular algorithm or method varies over different systems or devices \cite{li2020fair}, \cite{wang2021federated}. A given system is considered fairer than another system if its AV is less than the other system.
\cite{lyu2020collaborative} used the Pearson correlation coefficient for the quantification of collaboration fairness where the fairness range lies within [-1,1], with larger values signifying more fairness.
 \begin{equation}
 C_{a b}=\frac{\sum_{i=1}^n\left(a_i-\bar{a}\right)\left(b_i-\bar{b}\right)}{(N-1) std_a std_b}
 \end{equation}
A negative coefficient, on the other hand, indicates unequal treatment. Here, $\bar{a}$ and $\bar{b}$ are the sample means of a and b, respectively, while $std_{a}$ and $std_{b}$ are the corrected standard deviations.

Different distance metrics, for instance, Euclidean distance, Minkowski Distance and Manhattan distance, etc are also used to measure fairness between systems by evaluating the similarity/dissimilarity between their performance across different systems and for client contribution evaluation \cite{9006327}, \cite{liu2021gtgshapley}. 

\section{Preserving Model Fairness, Accuracy and Privacy}
Machine learning applications are expanding exponentially, and with that growth comes a greater requirement to guarantee model fairness and accuracy while safeguarding user data privacy. These applications, however, may display unexpected behaviors, such as unfairness, which cause groups with certain sensitive features (for instance, race), to experience varying patterns of outcomes. It is important for ML systems to strike a reasonable balance between model fairness, privacy, and accuracy. Table 4 states some works that try to achieve both fairness and privacy. 

FGFL \cite{LI2023968} evaluates players based on reputation and contribution factors and produces a blockchain-based incentive governor for Federated Learning. The task publisher equitably pays clients in order to recruit efficient ones, while malicious ones are penalized and discarded. \cite{hu2022fair} develops the concept of Bounded Group Loss as a theoretically informed approach to group fairness and provides a scalable federated optimization approach that optimizes the empirical risk under a number of group fairness requirements using this configuration. Based on a dynamic real-time worker evaluation process, FIFL \cite{10.1145/3472456.3472469} equitably compensates workers in order to attract trustworthy and efficient workers while punishing and removing malevolent ones.

\section{Challenges and Future Directions in Fairness and Privacy Preservation in FL}
\textbf{Lack of Metric to Measure Effort Costs in Contribution Evaluation.} There lies a transient discrepancy between the client contributions and rewards in FL because no notable metric exists to measure the temporal costs such as waiting time, cost of efforts, etc. Most contribution assessment studies in FL systems concentrate on assessing data collecting costs, data quality, model performance, total revenue generated, etc. For absolute fair treatment among clients, the temporal costs such as waiting times for getting the complete payout must be taken into consideration when the incentive for FL participants depends on future revenues generated. \\
\textbf{Missing or Limited Sensitive Features.}
In reality, sensitive qualities may not be disclosed or readily available, despite that most studies currently assume that that information is included in the dataset. Much of the sensitive attributes may be missing or limited due to limitations set by some regulations. In this scenario, it is required to achieve a balance between group fairness, individual fairness, and privacy despite lacking sensitive features explicitly. \\
\textbf{Issues in Vertical FL.}
Even though numerous privacy-preserving procedures, such as DP, secure multiparty computation (SMC), homomorphic encryption (HE), and their hybrid variations, have been widely employed in HFL, very few of these techniques have been investigated in VFL. \cite{wei2022vertical} stated that the difficulty for VFL is the possession of secret properties or outputs by specific models caused by the splitting techniques. These properties tend to have an impact on the level of security and privacy of a model in FL.
The splitting design is easy for basic ML models like Logistic Regression (LR) and kernel models, but privacy-preserving techniques are quite difficult since the input is processed in an insecure linear way. Less private information will be exposed in the intermediate outputs if complicated neural networks, such as the Convolutional Neural Network, are used as the bottom model. This is due to the possibility of improving dispersed data privacy using a complicated nonlinear function. An important difficulty in privacy-preserving VFL systems is comprehending and causing adjustments between privacy, operational efficiency, and performance. \\
\textbf{Privacy Heterogeneity Among Devices.} Most current privacy-preserving FL systems take into account that all devices have the same privacy restrictions and employ the same techniques to protect model weights or gradients. Few works consider the notion of privacy heterogeneity in FL systems, which revolves around the idea that each party's data cannot be used to its fullest potential provided that each party has different privacy restrictions. Systems that handle devices differently based on privacy needs should be established, and alternative measures should be taken to secure model parameters accordingly, to boost the learning capabilities of a model. \\
\textbf{Limitations of Non-Monopolistic FL settings.} Most existing works in Federated learning are based on a single server-multiple client scheme, resembling a monopolistic market. The server motivates the clients to take part in the federated model training and distributes incentives. Under this premise, only one server assigns tasks and this system does not promote parallel task execution from a server's point of view. All the clients in these types of monopolistic systems are maintained by a single server and the clients have no choice to choose servers. They only have the option to choose whether or not to participate. Such a system hinders the advancement of fairness-conscious FL systems. A non-monopolistic system consisting of multiple servers involves parallel task assignment, global model updation, and execution. They can compete with one another to acquire more customers and more clients can gain the willingness to participate provided that they can choose their desired server. This also causes increased pressure to allocate more fair incentives, causing clients to gain motivation to join the scheme.\\
\textbf{Computational Challenges in Privacy-Preserving Fairness-Aware FL.} Privacy-preserving fairness-aware FL systems are increasingly becoming computationally expensive and large models are particularly vulnerable to privacy assaults and fairness issues because they have high-dimensional parameter vectors and numerous sensitive attributes. This problem should be solved by facilitating local updating, curse of dimensionality reduction, and model compression. Training Federated networks on limited resources and low-power devices present unique obstacles such as massive computation, storage inadequacy, and battery limitations.  Trusted execution environments(TEE) setups and maintenance are extremely costly and difficult. Other hardware-based encryption solutions should also be considered to protect data security and boost encryption efficiency, such as combining FL with new protection technologies such as blockchain to assure FL cross-domain learning security. A limited number of works have used such a method.  It is crucial to do a thorough analysis of the trade-offs between accuracy, privacy, and communication for each strategy and facilitate inexpensive and easy parallel processing and TEE setup.\\
\textbf{Need for Continual Learning Methods.} Training a machine learning model is a costly and time-consuming job, which can be exacerbated in a federated learning environment. A trained model's performance degrades when data distributions change over time and when the available data per client is negligible or absent. To improve federated model learning capabilities, methods such as One-Shot Knowledge  Distillation \cite{zhang2021practical},  Data-Free knowledge Distillation \cite{zhu2021data} or a combination  \cite{zhang2022dense}  exist insufficiently, but should be made more common. To reduce the high cost of federated training, it is compulsory to investigate ways for improving federated model learning capacity over time. In that case, a limited amount of work exists concerning Meta-learning \cite{criado2022non}, \cite{fallah2020personalized}, \cite{jiang2019improving}, online learning \cite{han2020adaptive}, and continual learning \cite{usmanova2021distillation}, \cite{criado2022non} which will not impede a model's performance and learning over time.\\
\textbf{Tradeoff Between Privacy, Accuracy, and Fairness in FL.} At present, it has become increasingly difficult to strike a perfect balance between privacy, fairness, and accuracy in federated learning systems. \\
\textbf{Appropriate Privacy Preserving and Fairness Evaluation Metric.} Different systems require different privacy-preserving and fairness evaluation metrics and it is imperative to find the correct one that suits a particular situation since each metric might yield a different outcome.\\
\textbf{Choosing the Right Privacy-Preserving Method.} Since each privacy-preserving technique possesses its own kind of advantage and disadvantage as discussed in Table 1, it is required to consider an aggregation of multiple techniques that benefits a particular scenario and yields the best results.
\section{Conclusion}
This paper provides a summary of recent research on fairness and privacy concerns in FL, in addition to insights from earlier studies. We have also made an effort to provide some insights on challenges and issues related to the implementation of fairness-aware and privacy-preserving FL systems. Our work is intended to assist researchers in the field of FL in evaluating the current state of research on growing privacy and fairness-related concerns. We believe that AI practitioners and system designers will also benefit from this survey, as it provides a concise and understandable introduction to the topic. It is expected that significant efforts will be invested in FL research in the future for its potential as an alternative to centralized ML approaches in a variety of privacy and resource-critical scenarios. Privacy threats and fairness-related issues will continue to grow as the field of FL grows. We hope that significant efforts will be made in the future to overcome these obstacles, and we believe that our survey, the first of its kind to combine privacy and fairness concerns related to FL, will make a significant contribution in this regard.

\ifCLASSOPTIONcompsoc

  \section*{Acknowledgments}
\else

  \section*{Acknowledgment}
\fi
This work was partly supported by (1) Institute of Information \& communications Technology Planning \& Evaluation (IITP) grant funded by the Korea government (MSIT) (No.2020-0-01373, Artificial Intelligence Graduate School Program (Hanyang University)) and (2) the Bio \& Medical Technology Development Program of the National Research Foundation (NRF) funded by the Korean government (MSIT) (No. NRF-2021M3E5D2A01021156).

{\small
\bibliographystyle{ieee_fullname}
\bibliography{new}

\begin{thebibliography}{100}\itemsep=-1pt

\bibitem{abay2020mitigating}
Annie Abay, Yi Zhou, Nathalie Baracaldo, Shashank Rajamoni, Ebube Chuba, and
  Heiko Ludwig.
\newblock Mitigating bias in federated learning, 2020.

\bibitem{abdulrahman2020survey}
Sawsan AbdulRahman, Hanine Tout, Hakima Ould-Slimane, Azzam Mourad, Chamseddine
  Talhi, and Mohsen Guizani.
\newblock A survey on federated learning: The journey from centralized to
  distributed on-site learning and beyond.
\newblock {\em IEEE Internet of Things Journal}, 8(7):5476--5497, 2020.

\bibitem{adnan2022federated}
Mohammed Adnan, Shivam Kalra, Jesse~C Cresswell, Graham~W Taylor, and Hamid~R
  Tizhoosh.
\newblock Federated learning and differential privacy for medical image
  analysis.
\newblock {\em Scientific reports}, 12(1):1953, 2022.

\bibitem{10.1093/comjnl/bxab025}
Waqar Ali, Rajesh Kumar, Zhiyi Deng, Yansong Wang, and Jie Shao.
\newblock {A Federated Learning Approach for Privacy Protection in
  Context-Aware Recommender Systems}.
\newblock {\em The Computer Journal}, 64(7):1016--1027, 04 2021.

\bibitem{9833842}
Saba Amiri, Adam Belloum, Eric Nalisnick, Sander Klous, and Leon Gommans.
\newblock On the impact of non-iid data on the performance and fairness of
  differentially private federated learning.
\newblock In {\em 2022 52nd Annual IEEE/IFIP International Conference on
  Dependable Systems and Networks Workshops (DSN-W)}, pages 52--58, 2022.

\bibitem{angulo2022synthetic}
Cecilio Angulo and Crist{\'o}bal Raya.
\newblock Synthetic data for anonymization in secure data spaces for federated
  learning.
\newblock In {\em Artificial Intelligence Research and Development}, pages
  91--94. IOS Press, 2022.

\bibitem{awasthi2020equalized}
Pranjal Awasthi, Matthäus Kleindessner, and Jamie Morgenstern.
\newblock Equalized odds postprocessing under imperfect group information,
  2020.

\bibitem{bartlett2008classification}
Peter~L Bartlett and Marten~H Wegkamp.
\newblock Classification with a reject option using a hinge loss.
\newblock {\em Journal of Machine Learning Research}, 9(8), 2008.

\bibitem{basu2021benchmarking}
Priyam Basu, Tiasa~Singha Roy, Rakshit Naidu, Zumrut Muftuoglu, Sahib Singh,
  and Fatemehsadat Mireshghallah.
\newblock Benchmarking differential privacy and federated learning for bert
  models.
\newblock {\em arXiv preprint arXiv:2106.13973}, 2021.

\bibitem{bayatbabolghani2018secure}
Fattaneh Bayatbabolghani and Marina Blanton.
\newblock Secure multi-party computation.
\newblock In {\em Proceedings of the 2018 ACM SIGSAC conference on computer and
  communications security}, pages 2157--2159, 2018.

\bibitem{beimel2011secret}
Amos Beimel.
\newblock Secret-sharing schemes: A survey.
\newblock In {\em Coding and Cryptology: Third International Workshop, IWCC
  2011, Qingdao, China, May 30-June 3, 2011. Proceedings 3}, pages 11--46.
  Springer, 2011.

\bibitem{bettini2006role}
Claudio Bettini, X.~Sean Wang, and Sushil Jajodia.
\newblock The role of quasi-identifiers in k-anonymity revisited, 2006.

\bibitem{bhowmick2019protection}
Abhishek Bhowmick, John Duchi, Julien Freudiger, Gaurav Kapoor, and Ryan
  Rogers.
\newblock Protection against reconstruction and its applications in private
  federated learning, 2019.

\bibitem{10.1145/3029806.3029829}
Vincent Bindschaedler, Shantanu Rane, Alejandro~E. Brito, Vanishree Rao, and
  Ersin Uzun.
\newblock Achieving differential privacy in secure multiparty data aggregation
  protocols on star networks.
\newblock In {\em Proceedings of the Seventh ACM on Conference on Data and
  Application Security and Privacy}, CODASPY '17, page 115–125, New York, NY,
  USA, 2017. Association for Computing Machinery.

\bibitem{biswas2021fair}
Sumon Biswas and Hridesh Rajan.
\newblock Fair preprocessing: towards understanding compositional fairness of
  data transformers in machine learning pipeline.
\newblock In {\em Proceedings of the 29th ACM Joint Meeting on European
  Software Engineering Conference and Symposium on the Foundations of Software
  Engineering}, pages 981--993, 2021.

\bibitem{bonawitz2016practical}
Keith Bonawitz, Vladimir Ivanov, Ben Kreuter, Antonio Marcedone, H.~Brendan
  McMahan, Sarvar Patel, Daniel Ramage, Aaron Segal, and Karn Seth.
\newblock Practical secure aggregation for federated learning on user-held
  data, 2016.

\bibitem{calmon2017optimized}
Flavio~P. Calmon, Dennis Wei, Karthikeyan~Natesan Ramamurthy, and Kush~R.
  Varshney.
\newblock Optimized data pre-processing for discrimination prevention, 2017.

\bibitem{8975792}
Di Cao, Shan Chang, Zhijian Lin, Guohua Liu, and Donghong Sun.
\newblock Understanding distributed poisoning attack in federated learning.
\newblock In {\em 2019 IEEE 25th International Conference on Parallel and
  Distributed Systems (ICPADS)}, pages 233--239, 2019.

\bibitem{cao2020ifed}
Hui Cao, Shubo Liu, Renfang Zhao, and Xingxing Xiong.
\newblock Ifed: A novel federated learning framework for local differential
  privacy in power internet of things.
\newblock {\em International Journal of Distributed Sensor Networks},
  16(5):1550147720919698, 2020.

\bibitem{10020554}
Alycia~N. Carey, Wei Du, and Xintao Wu.
\newblock Robust personalized federated learning under demographic fairness
  heterogeneity.
\newblock In {\em 2022 IEEE International Conference on Big Data (Big Data)},
  pages 1425--1434, 2022.

\bibitem{Chakraborty_2020}
Joymallya Chakraborty, Kewen Peng, and Tim Menzies.
\newblock Making fair {ML} software using trustworthy explanation.
\newblock In {\em Proceedings of the 35th {IEEE}/{ACM} International Conference
  on Automated Software Engineering}. {ACM}, dec 2020.

\bibitem{chamani2020mitigating}
Javad~Ghareh Chamani and Dimitrios Papadopoulos.
\newblock Mitigating leakage in federated learning with trusted hardware, 2020.

\bibitem{Chamikara_2021}
M.A.P. Chamikara, P. Bertok, I. Khalil, D. Liu, and S. Camtepe.
\newblock Privacy preserving distributed machine learning with federated
  learning.
\newblock {\em Computer Communications}, 171:112--125, apr 2021.

\bibitem{chang2019cronus}
Hongyan Chang, Virat Shejwalkar, Reza Shokri, and Amir Houmansadr.
\newblock Cronus: Robust and heterogeneous collaborative learning with
  black-box knowledge transfer, 2019.

\bibitem{chen2008survey}
Keke Chen and Ling Liu.
\newblock A survey of multiplicative perturbation for privacy-preserving data
  mining.
\newblock {\em Privacy-preserving data mining: models and algorithms}, pages
  157--181, 2008.

\bibitem{chen2022decentralized}
Shuzhen Chen, Dongxiao Yu, Yifei Zou, Jiguo Yu, and Xiuzhen Cheng.
\newblock Decentralized wireless federated learning with differential privacy.
\newblock {\em IEEE Transactions on Industrial Informatics}, 18(9):6273--6282,
  2022.

\bibitem{chen2022federated}
Yao Chen, Yijie Gui, Hong Lin, Wensheng Gan, and Yongdong Wu.
\newblock Federated learning attacks and defenses: A survey, 2022.

\bibitem{chen2020training}
Yu Chen, Fang Luo, Tong Li, Tao Xiang, Zheli Liu, and Jin Li.
\newblock A training-integrity privacy-preserving federated learning scheme
  with trusted execution environment.
\newblock {\em Information Sciences}, 522:69--79, 2020.

\bibitem{cheng2021secureboost}
Kewei Cheng, Tao Fan, Yilun Jin, Yang Liu, Tianjian Chen, Dimitrios
  Papadopoulos, and Qiang Yang.
\newblock Secureboost: A lossless federated learning framework, 2021.

\bibitem{cho2020client}
Yae~Jee Cho, Jianyu Wang, and Gauri Joshi.
\newblock Client selection in federated learning: Convergence analysis and
  power-of-choice selection strategies, 2020.

\bibitem{choquettechoo2021capc}
Christopher~A. Choquette-Choo, Natalie Dullerud, Adam Dziedzic, Yunxiang Zhang,
  Somesh Jha, Nicolas Papernot, and Xiao Wang.
\newblock Capc learning: Confidential and private collaborative learning, 2021.

\bibitem{choudhury2019differential}
Olivia Choudhury, Aris Gkoulalas-Divanis, Theodoros Salonidis, Issa Sylla,
  Yoonyoung Park, Grace Hsu, and Amar Das.
\newblock Differential privacy-enabled federated learning for sensitive health
  data.
\newblock {\em arXiv preprint arXiv:1910.02578}, 2019.

\bibitem{choudhury2020anonymizing}
Olivia Choudhury, Aris Gkoulalas-Divanis, Theodoros Salonidis, Issa Sylla,
  Yoonyoung Park, Grace Hsu, and Amar Das.
\newblock Anonymizing data for privacy-preserving federated learning, 2020.

\bibitem{choudhury2020syntactic}
Olivia Choudhury, Aris Gkoulalas-Divanis, Theodoros Salonidis, Issa Sylla,
  Yoonyoung Park, Grace Hsu, and Amar Das.
\newblock A syntactic approach for privacy-preserving federated learning.
\newblock In {\em ECAI 2020}, pages 1762--1769. IOS Press, 2020.

\bibitem{criado2022non}
Marcos~F Criado, Fernando~E Casado, Roberto Iglesias, Carlos~V Regueiro, and
  Sen{\'e}n Barro.
\newblock Non-iid data and continual learning processes in federated learning:
  A long road ahead.
\newblock {\em Information Fusion}, 88:263--280, 2022.

\bibitem{darlow2020latent}
Luke Darlow, Stanislaw Jastrzebski, and Amos Storkey.
\newblock Latent adversarial debiasing: Mitigating collider bias in deep neural
  networks.
\newblock {\em arXiv preprint arXiv:2011.11486}, 2020.

\bibitem{dwork2012fairness}
Cynthia Dwork, Moritz Hardt, Toniann Pitassi, Omer Reingold, and Richard Zemel.
\newblock Fairness through awareness.
\newblock In {\em Proceedings of the 3rd innovations in theoretical computer
  science conference}, pages 214--226, 2012.

\bibitem{ezzeldin2022fairfed}
Yahya~H. Ezzeldin, Shen Yan, Chaoyang He, Emilio Ferrara, and Salman
  Avestimehr.
\newblock Fairfed: Enabling group fairness in federated learning, 2022.

\bibitem{fallah2020personalized}
Alireza Fallah, Aryan Mokhtari, and Asuman Ozdaglar.
\newblock Personalized federated learning with theoretical guarantees: A
  model-agnostic meta-learning approach.
\newblock {\em Advances in Neural Information Processing Systems},
  33:3557--3568, 2020.

\bibitem{fan2012somewhat}
Junfeng Fan and Frederik Vercauteren.
\newblock Somewhat practical fully homomorphic encryption.
\newblock {\em Cryptology ePrint Archive}, 2012.

\bibitem{fan2022fair}
Zhenan Fan, Huang Fang, Zirui Zhou, Jian Pei, Michael~P Friedlander, and Yong
  Zhang.
\newblock Fair and efficient contribution valuation for vertical federated
  learning.
\newblock {\em arXiv preprint arXiv:2201.02658}, 2022.

\bibitem{friedler2019comparative}
Sorelle~A Friedler, Carlos Scheidegger, Suresh Venkatasubramanian, Sonam
  Choudhary, Evan~P Hamilton, and Derek Roth.
\newblock A comparative study of fairness-enhancing interventions in machine
  learning.
\newblock In {\em Proceedings of the conference on fairness, accountability,
  and transparency}, pages 329--338, 2019.

\bibitem{fung2019dancing}
Clement Fung, Jamie Koerner, Stewart Grant, and Ivan Beschastnikh.
\newblock Dancing in the dark: Private multi-party machine learning in an
  untrusted setting, 2019.

\bibitem{8619133}
Shripad Gade and Nitin~H. Vaidya.
\newblock Privacy-preserving distributed learning via obfuscated stochastic
  gradients.
\newblock In {\em 2018 IEEE Conference on Decision and Control (CDC)}, pages
  184--191, 2018.

\bibitem{Galli_2023}
Filippo Galli, Sayan Biswas, Kangsoo Jung, Tommaso Cucinotta, and Catuscia
  Palamidessi.
\newblock Group privacy for personalized federated learning.
\newblock In {\em Proceedings of the 9th International Conference on
  Information Systems Security and Privacy}. {SCITEPRESS} - Science and
  Technology Publications, 2023.

\bibitem{9005992}
Dashan Gao, Yang Liu, Anbu Huang, Ce Ju, Han Yu, and Qiang Yang.
\newblock Privacy-preserving heterogeneous federated transfer learning.
\newblock In {\em 2019 IEEE International Conference on Big Data (Big Data)},
  pages 2552--2559, 2019.

\bibitem{10.1145/3472456.3472469}
Liang Gao, Li Li, Yingwen Chen, Wenli Zheng, ChengZhong Xu, and Ming Xu.
\newblock Fifl: A fair incentive mechanism for federated learning.
\newblock In {\em Proceedings of the 50th International Conference on Parallel
  Processing}, ICPP '21, New York, NY, USA, 2021. Association for Computing
  Machinery.

\bibitem{geyer2018differentially}
Robin~C. Geyer, Tassilo Klein, and Moin Nabi.
\newblock Differentially private federated learning: A client level
  perspective, 2018.

\bibitem{ghorbani2019data}
Amirata Ghorbani and James Zou.
\newblock Data shapley: Equitable valuation of data for machine learning, 2019.

\bibitem{girgis2021shuffled}
Antonious Girgis, Deepesh Data, Suhas Diggavi, Peter Kairouz, and
  Ananda~Theertha Suresh.
\newblock Shuffled model of differential privacy in federated learning.
\newblock In {\em International Conference on Artificial Intelligence and
  Statistics}, pages 2521--2529. PMLR, 2021.

\bibitem{han2020adaptive}
Pengchao Han, Shiqiang Wang, and Kin~K Leung.
\newblock Adaptive gradient sparsification for efficient federated learning: An
  online learning approach.
\newblock In {\em 2020 IEEE 40th international conference on distributed
  computing systems (ICDCS)}, pages 300--310. IEEE, 2020.

\bibitem{9770028}
Weituo Hao, Nikhil Mehta, Kevin~J. Liang, Pengyu Cheng, Mostafa El-Khamy, and
  Lawrence Carin.
\newblock Waffle: Weight anonymized factorization for federated learning.
\newblock {\em IEEE Access}, 10:49207--49218, 2022.

\bibitem{hardt2016equality}
Moritz Hardt, Eric Price, and Nathan Srebro.
\newblock Equality of opportunity in supervised learning, 2016.

\bibitem{herbei2006classification}
Radu Herbei and Marten~H Wegkamp.
\newblock Classification with reject option.
\newblock {\em The Canadian Journal of Statistics/La Revue Canadienne de
  Statistique}, pages 709--721, 2006.

\bibitem{hitaj2017deep}
Briland Hitaj, Giuseppe Ateniese, and Fernando Perez-Cruz.
\newblock Deep models under the gan: Information leakage from collaborative
  deep learning, 2017.

\bibitem{hong2021federated}
Junyuan Hong, Zhuangdi Zhu, Shuyang Yu, Zhangyang Wang, Hiroko~H Dodge, and
  Jiayu Zhou.
\newblock Federated adversarial debiasing for fair and transferable
  representations.
\newblock In {\em Proceedings of the 27th ACM SIGKDD Conference on Knowledge
  Discovery \& Data Mining}, pages 617--627, 2021.

\bibitem{horvath2022fjord}
Samuel Horvath, Stefanos Laskaridis, Mario Almeida, Ilias Leontiadis,
  Stylianos~I. Venieris, and Nicholas~D. Lane.
\newblock Fjord: Fair and accurate federated learning under heterogeneous
  targets with ordered dropout, 2022.

\bibitem{9679121}
Hongsheng Hu, Zoran Salcic, Lichao Sun, Gillian Dobbie, and Xuyun Zhang.
\newblock Source inference attacks in federated learning.
\newblock In {\em 2021 IEEE International Conference on Data Mining (ICDM)},
  pages 1102--1107, 2021.

\bibitem{hu2020personalized}
Rui Hu, Yuanxiong Guo, Hongning Li, Qingqi Pei, and Yanmin Gong.
\newblock Personalized federated learning with differential privacy.
\newblock {\em IEEE Internet of Things Journal}, 7(10):9530--9539, 2020.

\bibitem{Hu_2022}
Sixu Hu, Yuan Li, Xu Liu, Qinbin Li, Zhaomin Wu, and Bingsheng He.
\newblock The {OARF} benchmark suite: Characterization and implications for
  federated learning systems.
\newblock {\em {ACM} Transactions on Intelligent Systems and Technology},
  13(4):1--32, jun 2022.

\bibitem{hu2022fair}
Shengyuan Hu, Zhiwei~Steven Wu, and Virginia Smith.
\newblock Fair federated learning via bounded group loss, 2022.

\bibitem{hu2023federated}
Zeou Hu, Kiarash Shaloudegi, Guojun Zhang, and Yaoliang Yu.
\newblock Federated learning meets multi-objective optimization, 2023.

\bibitem{huang2022aflpc}
Jie Huang, Cheng Xu, Zhaohua Ji, Shan Xiao, Teng Liu, Nan Ma, and Qinghui Zhou.
\newblock Aflpc: an asynchronous federated learning privacy-preserving
  computing model applied to 5g-v2x.
\newblock {\em Security and Communication Networks}, 2022, 2022.

\bibitem{huang2022stochastic}
Tiansheng Huang, Weiwei Lin, Li Shen, Keqin Li, and Albert~Y. Zomaya.
\newblock Stochastic client selection for federated learning with volatile
  clients, 2022.

\bibitem{Huang_2020}
Tiansheng Huang, Weiwei Lin, Wentai Wu, Ligang He, Keqin Li, and Albert Zomaya.
\newblock An efficiency-boosting client selection scheme for federated learning
  with fairness guarantee.
\newblock {\em {IEEE} Transactions on Parallel and Distributed Systems}, pages
  1--1, 2020.

\bibitem{9272649}
Tiansheng Huang, Weiwei Lin, Wentai Wu, Ligang He, Keqin Li, and Albert~Y.
  Zomaya.
\newblock An efficiency-boosting client selection scheme for federated learning
  with fairness guarantee.
\newblock {\em IEEE Transactions on Parallel and Distributed Systems},
  32(7):1552--1564, 2021.

\bibitem{huang2020fairness}
Wei Huang, Tianrui Li, Dexian Wang, Shengdong Du, and Junbo Zhang.
\newblock Fairness and accuracy in federated learning, 2020.

\bibitem{huang2021evaluating}
Yangsibo Huang, Samyak Gupta, Zhao Song, Kai Li, and Sanjeev Arora.
\newblock Evaluating gradient inversion attacks and defenses in federated
  learning.
\newblock {\em Advances in Neural Information Processing Systems},
  34:7232--7241, 2021.

\bibitem{9165379}
Selim Ickin, Konstantinos Vandikas, Farnaz Moradi, Jalil Taghia, and Wenfeng
  Hu.
\newblock Ensemble-based synthetic data synthesis for federated qoe modeling.
\newblock In {\em 2020 6th IEEE Conference on Network Softwarization
  (NetSoft)}, pages 72--76, 2020.

\bibitem{9443523}
Yae Jee~Cho, Samarth Gupta, Gauri Joshi, and Osman Yağan.
\newblock Bandit-based communication-efficient client selection strategies for
  federated learning.
\newblock In {\em 2020 54th Asilomar Conference on Signals, Systems, and
  Computers}, pages 1066--1069, 2020.

\bibitem{jia2021blockchain}
Bin Jia, Xiaosong Zhang, Jiewen Liu, Yang Zhang, Ke Huang, and Yongquan Liang.
\newblock Blockchain-enabled federated learning data protection aggregation
  scheme with differential privacy and homomorphic encryption in iiot.
\newblock {\em IEEE Transactions on Industrial Informatics}, 18(6):4049--4058,
  2021.

\bibitem{9448383}
Bin Jia, Xiaosong Zhang, Jiewen Liu, Yang Zhang, Ke Huang, and Yongquan Liang.
\newblock Blockchain-enabled federated learning data protection aggregation
  scheme with differential privacy and homomorphic encryption in iiot.
\newblock {\em IEEE Transactions on Industrial Informatics}, 18(6):4049--4058,
  2022.

\bibitem{jiang2021privacy}
Bin Jiang, Jianqiang Li, Huihui Wang, and Houbing Song.
\newblock Privacy-preserving federated learning for industrial edge computing
  via hybrid differential privacy and adaptive compression.
\newblock {\em IEEE Transactions on Industrial Informatics}, 19(2):1136--1144,
  2021.

\bibitem{jiang2022comprehensive}
Xue Jiang, Xuebing Zhou, and Jens Grossklags.
\newblock Comprehensive analysis of privacy leakage in vertical federated
  learning during prediction.
\newblock {\em Proc. Priv. Enhancing Technol.}, 2022(2):263--281, 2022.

\bibitem{jiang2019improving}
Yihan Jiang, Jakub Kone{\v{c}}n{\`y}, Keith Rush, and Sreeram Kannan.
\newblock Improving federated learning personalization via model agnostic meta
  learning.
\newblock {\em arXiv preprint arXiv:1909.12488}, 2019.

\bibitem{jiang2021flashe}
Zhifeng Jiang, Wei Wang, and Yang Liu.
\newblock Flashe: Additively symmetric homomorphic encryption for cross-silo
  federated learning, 2021.

\bibitem{juarez2023you}
Marc Juarez and Aleksandra Korolova.
\newblock "you can't fix what you can't measure": Privately measuring
  demographic performance disparities in federated learning, 2023.

\bibitem{10.1561/2200000083}
Peter Kairouz, H.~Brendan McMahan, Brendan Avent, Aur\'{e}lien Bellet, Mehdi
  Bennis, Arjun Nitin~Bhagoji, Kallista Bonawitz, Zachary Charles, Graham
  Cormode, Rachel Cummings, Rafael G.~L. D’Oliveira, Hubert Eichner, Salim
  El~Rouayheb, David Evans, Josh Gardner, Zachary Garrett, Adri\`{a}
  Gasc\'{o}n, Badih Ghazi, Phillip~B. Gibbons, Marco Gruteser, Zaid Harchaoui,
  Chaoyang He, Lie He, Zhouyuan Huo, Ben Hutchinson, Justin Hsu, Martin Jaggi,
  Tara Javidi, Gauri Joshi, Mikhail Khodak, Jakub Konecn\'{y}, Aleksandra
  Korolova, Farinaz Koushanfar, Sanmi Koyejo, Tancr\`{e}de Lepoint, Yang Liu,
  Prateek Mittal, Mehryar Mohri, Richard Nock, Ayfer \"{O}zg\"{u}r, Rasmus
  Pagh, Hang Qi, Daniel Ramage, Ramesh Raskar, Mariana Raykova, Dawn Song,
  Weikang Song, Sebastian~U. Stich, Ziteng Sun, Ananda~Theertha Suresh, Florian
  Tram\`{e}r, Praneeth Vepakomma, Jianyu Wang, Li Xiong, Zheng Xu, Qiang Yang,
  Felix~X. Yu, Han Yu, and Sen Zhao.
\newblock Advances and open problems in federated learning.
\newblock {\em Found. Trends Mach. Learn.}, 14(1–2):1–210, jun 2021.

\bibitem{kamishima2012fairness}
Toshihiro Kamishima, Shotaro Akaho, Hideki Asoh, and Jun Sakuma.
\newblock Fairness-aware classifier with prejudice remover regularizer.
\newblock In {\em Machine Learning and Knowledge Discovery in Databases:
  European Conference, ECML PKDD 2012, Bristol, UK, September 24-28, 2012.
  Proceedings, Part II 23}, pages 35--50. Springer, 2012.

\bibitem{kamishima2011fairness}
Toshihiro Kamishima, Shotaro Akaho, and Jun Sakuma.
\newblock Fairness-aware learning through regularization approach.
\newblock In {\em 2011 IEEE 11th International Conference on Data Mining
  Workshops}, pages 643--650. IEEE, 2011.

\bibitem{kanaparthy2022f3}
Samhita Kanaparthy, Manisha Padala, Sankarshan Damle, Ravi~Kiran
  Sarvadevabhatla, and Sujit Gujar.
\newblock F3: Fair and federated face attribute classification with
  heterogeneous data, 2022.

\bibitem{kang2019incentive}
Jiawen Kang, Zehui Xiong, Dusit Niyato, Han Yu, Ying-Chang Liang, and Dong~In
  Kim.
\newblock Incentive design for efficient federated learning in mobile networks:
  A contract theory approach, 2019.

\bibitem{karimireddy2021scaffold}
Sai~Praneeth Karimireddy, Satyen Kale, Mehryar Mohri, Sashank~J. Reddi,
  Sebastian~U. Stich, and Ananda~Theertha Suresh.
\newblock Scaffold: Stochastic controlled averaging for federated learning,
  2021.

\bibitem{4690986}
Shiva~Prasad Kasiviswanathan, Homin~K. Lee, Kobbi Nissim, Sofya Raskhodnikova,
  and Adam Smith.
\newblock What can we learn privately?
\newblock In {\em 2008 49th Annual IEEE Symposium on Foundations of Computer
  Science}, pages 531--540, 2008.

\bibitem{9247530}
Latif~U. Khan, Shashi~Raj Pandey, Nguyen~H. Tran, Walid Saad, Zhu Han, Minh
  N.~H. Nguyen, and Choong~Seon Hong.
\newblock Federated learning for edge networks: Resource optimization and
  incentive mechanism.
\newblock {\em IEEE Communications Magazine}, 58(10):88--93, 2020.

\bibitem{khan2021federated}
Latif~U Khan, Walid Saad, Zhu Han, Ekram Hossain, and Choong~Seon Hong.
\newblock Federated learning for internet of things: Recent advances, taxonomy,
  and open challenges.
\newblock {\em IEEE Communications Surveys \& Tutorials}, 23(3):1759--1799,
  2021.

\bibitem{kiersztyn2021concept}
Adam Kiersztyn, Pawe{\l} Karczmarek, Krystyna Kiersztyn, Rafa{\l} {\L}opucki,
  Stanis{\l}aw Grzeg{\'o}rski, and Witold Pedrycz.
\newblock The concept of granular representation of the information potential
  of variables.
\newblock In {\em 2021 IEEE International Conference on Fuzzy Systems
  (FUZZ-IEEE)}, pages 1--6. IEEE, 2021.

\bibitem{kim2018logistic}
Andrey Kim, Yongsoo Song, Miran Kim, Keewoo Lee, and Jung~Hee Cheon.
\newblock Logistic regression model training based on the approximate
  homomorphic encryption.
\newblock {\em BMC medical genomics}, 11(4):23--31, 2018.

\bibitem{kim2021federated}
Muah Kim, Onur G{\"u}nl{\"u}, and Rafael~F Schaefer.
\newblock Federated learning with local differential privacy: Trade-offs
  between privacy, utility, and communication.
\newblock In {\em ICASSP 2021-2021 IEEE International Conference on Acoustics,
  Speech and Signal Processing (ICASSP)}, pages 2650--2654. IEEE, 2021.

\bibitem{kim2019multiaccuracy}
Michael~P Kim, Amirata Ghorbani, and James Zou.
\newblock Multiaccuracy: Black-box post-processing for fairness in
  classification.
\newblock In {\em Proceedings of the 2019 AAAI/ACM Conference on AI, Ethics,
  and Society}, pages 247--254, 2019.

\bibitem{9223632}
Sungwook Kim.
\newblock Incentive design and differential privacy based federated learning: A
  mechanism design perspective.
\newblock {\em IEEE Access}, 8:187317--187325, 2020.

\bibitem{9540342}
Prabhat Kumar, Govind~P. Gupta, and Rakesh Tripathi.
\newblock Pefl: Deep privacy-encoding-based federated learning framework for
  smart agriculture.
\newblock {\em IEEE Micro}, 42(1):33--40, 2022.

\bibitem{273723}
Fan Lai, Xiangfeng Zhu, Harsha~V. Madhyastha, and Mosharaf Chowdhury.
\newblock Oort: Efficient federated learning via guided participant selection.
\newblock In {\em 15th {USENIX} Symposium on Operating Systems Design and
  Implementation ({OSDI} 21)}, pages 19--35. {USENIX} Association, July 2021.

\bibitem{4221659}
Ninghui Li, Tiancheng Li, and Suresh Venkatasubramanian.
\newblock t-closeness: Privacy beyond k-anonymity and l-diversity.
\newblock In {\em 2007 IEEE 23rd International Conference on Data Engineering},
  pages 106--115, 2007.

\bibitem{li2021tilted}
Tian Li, Ahmad Beirami, Maziar Sanjabi, and Virginia Smith.
\newblock Tilted empirical risk minimization, 2021.

\bibitem{li2021ditto}
Tian Li, Shengyuan Hu, Ahmad Beirami, and Virginia Smith.
\newblock Ditto: Fair and robust federated learning through personalization,
  2021.

\bibitem{li2020fair}
Tian Li, Maziar Sanjabi, Ahmad Beirami, and Virginia Smith.
\newblock Fair resource allocation in federated learning, 2020.

\bibitem{LI2023968}
Xiaoli Li, Siran Zhao, Chuan Chen, and Zibin Zheng.
\newblock Heterogeneity-aware fair federated learning.
\newblock {\em Information Sciences}, 619:968--986, 2023.

\bibitem{li2019asynchronous}
Yanan Li, Shusen Yang, Xuebin Ren, and Cong Zhao.
\newblock Asynchronous federated learning with differential privacy for edge
  intelligence.
\newblock {\em arXiv preprint arXiv:1912.07902}, 2019.

\bibitem{li2020privacy}
Yong Li, Yipeng Zhou, Alireza Jolfaei, Dongjin Yu, Gaochao Xu, and Xi Zheng.
\newblock Privacy-preserving federated learning framework based on chained
  secure multiparty computing.
\newblock {\em IEEE Internet of Things Journal}, 8(8):6178--6186, 2020.

\bibitem{lian2021cofel}
Zhuotao Lian, Weizheng Wang, and Chunhua Su.
\newblock Cofel: Communication-efficient and optimized federated learning with
  local differential privacy.
\newblock In {\em ICC 2021-IEEE International Conference on Communications},
  pages 1--6. IEEE, 2021.

\bibitem{9929413}
Jialing Liao, Zheng Chen, and Erik~G. Larsson.
\newblock Over-the-air federated learning with privacy protection via
  correlated additive perturbations.
\newblock In {\em 2022 58th Annual Allerton Conference on Communication,
  Control, and Computing (Allerton)}, pages 1--8, 2022.

\bibitem{liu2022distributed}
Ji Liu, Jizhou Huang, Yang Zhou, Xuhong Li, Shilei Ji, Haoyi Xiong, and Dejing
  Dou.
\newblock From distributed machine learning to federated learning: A survey.
\newblock {\em Knowledge and Information Systems}, 64(4):885--917, 2022.

\bibitem{liu2008survey}
Kun Liu, Chris Giannella, and Hillol Kargupta.
\newblock A survey of attack techniques on privacy-preserving data perturbation
  methods.
\newblock {\em Privacy-Preserving Data Mining: Models and Algorithms}, pages
  359--381, 2008.

\bibitem{1549830}
Kun Liu, H. Kargupta, and J. Ryan.
\newblock Random projection-based multiplicative data perturbation for privacy
  preserving distributed data mining.
\newblock {\em IEEE Transactions on Knowledge and Data Engineering},
  18(1):92--106, 2006.

\bibitem{10.1145/3523227.3546771}
Shuchang Liu, Yingqiang Ge, Shuyuan Xu, Yongfeng Zhang, and Amelie Marian.
\newblock Fairness-aware federated matrix factorization.
\newblock In {\em Proceedings of the 16th ACM Conference on Recommender
  Systems}, RecSys '22, page 168–178, New York, NY, USA, 2022. Association
  for Computing Machinery.

\bibitem{LIU2022103309}
Tian Liu, Xueyang Hu, Hairuo Xu, Tao Shu, and Diep~N. Nguyen.
\newblock High-accuracy low-cost privacy-preserving federated learning in iot
  systems via adaptive perturbation.
\newblock {\em Journal of Information Security and Applications}, 70:103309,
  2022.

\bibitem{Liu_2020}
Yang Liu, Yan Kang, Chaoping Xing, Tianjian Chen, and Qiang Yang.
\newblock A secure federated transfer learning framework.
\newblock {\em {IEEE} Intelligent Systems}, 35(4):70--82, jul 2020.

\bibitem{Liu_2022}
Yang Liu, Yingting Liu, Zhijie Liu, Yuxuan Liang, Chuishi Meng, Junbo Zhang,
  and Yu Zheng.
\newblock Federated forest.
\newblock {\em {IEEE} Transactions on Big Data}, 8(3):843--854, jun 2022.

\bibitem{LIU202014}
Yang Liu, Zhuo Ma, Zheng Yan, Zhuzhu Wang, Ximeng Liu, and Jianfeng Ma.
\newblock Privacy-preserving federated k-means for proactive caching in next
  generation cellular networks.
\newblock {\em Information Sciences}, 521:14--31, 2020.

\bibitem{liu2022contract}
Yuan Liu, Mengmeng Tian, Yuxin Chen, Zehui Xiong, Cyril Leung, and Chunyan
  Miao.
\newblock A contract theory based incentive mechanism for federated learning.
\newblock In {\em Federated and Transfer Learning}, pages 117--137. Springer,
  2022.

\bibitem{liu2021gtgshapley}
Zelei Liu, Yuanyuan Chen, Han Yu, Yang Liu, and Lizhen Cui.
\newblock Gtg-shapley: Efficient and accurate participant contribution
  evaluation in federated learning, 2021.

\bibitem{liu2022privacypreserving}
Ziyao Liu, Jiale Guo, Wenzhuo Yang, Jiani Fan, Kwok-Yan Lam, and Jun Zhao.
\newblock Privacy-preserving aggregation in federated learning: A survey, 2022.

\bibitem{liu2019enhancing}
Zaoxing Liu, Tian Li, Virginia Smith, and Vyas Sekar.
\newblock Enhancing the privacy of federated learning with sketching.
\newblock {\em arXiv preprint arXiv:1911.01812}, 2019.

\bibitem{9686048}
Sin~Kit Lo, Yue Liu, Qinghua Lu, Chen Wang, Xiwei Xu, Hye-Young Paik, and
  Liming Zhu.
\newblock Toward trustworthy ai: Blockchain-based architecture design for
  accountability and fairness of federated learning systems.
\newblock {\em IEEE Internet of Things Journal}, 10(4):3276--3284, 2023.

\bibitem{lohia2019bias}
Pranay~K Lohia, Karthikeyan~Natesan Ramamurthy, Manish Bhide, Diptikalyan Saha,
  Kush~R Varshney, and Ruchir Puri.
\newblock Bias mitigation post-processing for individual and group fairness.
\newblock In {\em Icassp 2019-2019 ieee international conference on acoustics,
  speech and signal processing (icassp)}, pages 2847--2851. IEEE, 2019.

\bibitem{lyu2020collaborative}
Lingjuan Lyu, Xinyi Xu, and Qian Wang.
\newblock Collaborative fairness in federated learning, 2020.

\bibitem{lyu2022privacy}
Lingjuan Lyu, Han Yu, Xingjun Ma, Chen Chen, Lichao Sun, Jun Zhao, Qiang Yang,
  and Philip~S. Yu.
\newblock Privacy and robustness in federated learning: Attacks and defenses,
  2022.

\bibitem{lyu2020threats}
Lingjuan Lyu, Han Yu, and Qiang Yang.
\newblock Threats to federated learning: A survey, 2020.

\bibitem{ma2021privacypreserving}
Jing Ma, Si-Ahmed Naas, Stephan Sigg, and Xixiang Lyu.
\newblock Privacy-preserving federated learning based on multi-key homomorphic
  encryption, 2021.

\bibitem{pmlr-v162-marfoq22a}
Othmane Marfoq, Giovanni Neglia, Richard Vidal, and Laetitia Kameni.
\newblock Personalized federated learning through local memorization.
\newblock In Kamalika Chaudhuri, Stefanie Jegelka, Le Song, Csaba Szepesvari,
  Gang Niu, and Sivan Sabato, editors, {\em Proceedings of the 39th
  International Conference on Machine Learning}, volume 162 of {\em Proceedings
  of Machine Learning Research}, pages 15070--15092. PMLR, 17--23 Jul 2022.

\bibitem{mcmahan2018learning}
H.~Brendan McMahan, Daniel Ramage, Kunal Talwar, and Li Zhang.
\newblock Learning differentially private recurrent language models, 2018.

\bibitem{mishler2021fairness}
Alan Mishler, Edward~H Kennedy, and Alexandra Chouldechova.
\newblock Fairness in risk assessment instruments: Post-processing to achieve
  counterfactual equalized odds.
\newblock In {\em Proceedings of the 2021 ACM Conference on Fairness,
  Accountability, and Transparency}, pages 386--400, 2021.

\bibitem{7958569}
Payman Mohassel and Yupeng Zhang.
\newblock Secureml: A system for scalable privacy-preserving machine learning.
\newblock In {\em 2017 IEEE Symposium on Security and Privacy (SP)}, pages
  19--38, 2017.

\bibitem{pmlr-v97-mohri19a}
Mehryar Mohri, Gary Sivek, and Ananda~Theertha Suresh.
\newblock Agnostic federated learning.
\newblock In Kamalika Chaudhuri and Ruslan Salakhutdinov, editors, {\em
  Proceedings of the 36th International Conference on Machine Learning},
  volume~97 of {\em Proceedings of Machine Learning Research}, pages
  4615--4625. PMLR, 09--15 Jun 2019.

\bibitem{mohri2019agnostic}
Mehryar Mohri, Gary Sivek, and Ananda~Theertha Suresh.
\newblock Agnostic federated learning, 2019.

\bibitem{9581236}
Arup Mondal, Yash More, Ruthu~Hulikal Rooparaghunath, and Debayan Gupta.
\newblock Poster: Flatee: Federated learning across trusted execution
  environments.
\newblock In {\em 2021 IEEE European Symposium on Security and Privacy
  (EuroS\&P)}, pages 707--709, 2021.

\bibitem{mou2021verifiable}
Wenhao Mou, Chunlei Fu, Yan Lei, and Chunqiang Hu.
\newblock A verifiable federated learning scheme based on secure multi-party
  computation.
\newblock In {\em Wireless Algorithms, Systems, and Applications: 16th
  International Conference, WASA 2021, Nanjing, China, June 25--27, 2021,
  Proceedings, Part II}, pages 198--209. Springer, 2021.

\bibitem{mugunthan2019smpai}
Vaikkunth Mugunthan, Antigoni Polychroniadou, David Byrd, and Tucker~Hybinette
  Balch.
\newblock Smpai: Secure multi-party computation for federated learning.
\newblock In {\em Proceedings of the NeurIPS 2019 Workshop on Robust AI in
  Financial Services}, 2019.

\bibitem{munir2021fedprune}
Muhammad~Tahir Munir, Muhammad~Mustansar Saeed, Mahad Ali, Zafar~Ayyub Qazi,
  and Ihsan~Ayyub Qazi.
\newblock Fedprune: Towards inclusive federated learning, 2021.

\bibitem{muralidhar1999general}
Krishnamurty Muralidhar, Rahul Parsa, and Rathindra Sarathy.
\newblock A general additive data perturbation method for database security.
\newblock {\em management science}, 45(10):1399--1415, 1999.

\bibitem{nagalapatti2021game}
Lokesh Nagalapatti and Ramasuri Narayanam.
\newblock Game of gradients: Mitigating irrelevant clients in federated
  learning.
\newblock In {\em Proceedings of the AAAI Conference on Artificial
  Intelligence}, pages 9046--9054, 2021.

\bibitem{naseri2022local}
Mohammad Naseri, Jamie Hayes, and Emiliano~De Cristofaro.
\newblock Local and central differential privacy for robustness and privacy in
  federated learning, 2022.

\bibitem{nasr2019comprehensive}
Milad Nasr, Reza Shokri, and Amir Houmansadr.
\newblock Comprehensive privacy analysis of deep learning: Passive and active
  white-box inference attacks against centralized and federated learning.
\newblock In {\em 2019 IEEE symposium on security and privacy (SP)}, pages
  739--753. IEEE, 2019.

\bibitem{Nishio_2019}
Takayuki Nishio and Ryo Yonetani.
\newblock Client selection for federated learning with heterogeneous resources
  in mobile edge.
\newblock In {\em {ICC} 2019 - 2019 {IEEE} International Conference on
  Communications ({ICC})}. {IEEE}, may 2019.

\bibitem{ou2020homomorphic}
Wei Ou, Jianhuan Zeng, Zijun Guo, Wanqin Yan, Dingwan Liu, and Stelios Fuentes.
\newblock A homomorphic-encryption-based vertical federated learning scheme for
  rick management.
\newblock {\em Computer Science and Information Systems}, 17(3):819--834, 2020.

\bibitem{9714350}
Ahmed~El Ouadrhiri and Ahmed Abdelhadi.
\newblock Differential privacy for deep and federated learning: A survey.
\newblock {\em IEEE Access}, 10:22359--22380, 2022.

\bibitem{padala2021federated}
Manisha Padala, Sankarshan Damle, and Sujit Gujar.
\newblock Federated learning meets fairness and differential privacy, 2021.

\bibitem{paillier1999public}
Pascal Paillier.
\newblock Public-key cryptosystems based on composite degree residuosity
  classes.
\newblock In {\em Advances in Cryptology—EUROCRYPT’99: International
  Conference on the Theory and Application of Cryptographic Techniques Prague,
  Czech Republic, May 2--6, 1999 Proceedings 18}, pages 223--238. Springer,
  1999.

\bibitem{5288526}
Sinno~Jialin Pan and Qiang Yang.
\newblock A survey on transfer learning.
\newblock {\em IEEE Transactions on Knowledge and Data Engineering},
  22(10):1345--1359, 2010.

\bibitem{panda2022sparsefed}
Ashwinee Panda, Saeed Mahloujifar, Arjun~Nitin Bhagoji, Supriyo Chakraborty,
  and Prateek Mittal.
\newblock Sparsefed: Mitigating model poisoning attacks in federated learning
  with sparsification.
\newblock In {\em International Conference on Artificial Intelligence and
  Statistics}, pages 7587--7624. PMLR, 2022.

\bibitem{8995775}
Shashi~Raj Pandey, Nguyen~H. Tran, Mehdi Bennis, Yan~Kyaw Tun, Aunas Manzoor,
  and Choong~Seon Hong.
\newblock A crowdsourcing framework for on-device federated learning.
\newblock {\em IEEE Transactions on Wireless Communications}, 19(5):3241--3256,
  2020.

\bibitem{papadaki2021federating}
Afroditi Papadaki, Natalia Martinez, Martin Bertran, Guillermo Sapiro, and
  Miguel Rodrigues.
\newblock Federating for learning group fair models, 2021.

\bibitem{10.1145/3531146.3533081}
Afroditi Papadaki, Natalia Martinez, Martin Bertran, Guillermo Sapiro, and
  Miguel Rodrigues.
\newblock Minimax demographic group fairness in federated learning.
\newblock In {\em 2022 ACM Conference on Fairness, Accountability, and
  Transparency}, FAccT '22, page 142–159, New York, NY, USA, 2022.
  Association for Computing Machinery.

\bibitem{9952531}
Jaehyoung Park, Nam~Yul Yu, and Hyuk Lim.
\newblock Privacy-preserving federated learning using homomorphic encryption
  with different encryption keys.
\newblock In {\em 2022 13th International Conference on Information and
  Communication Technology Convergence (ICTC)}, pages 1869--1871, 2022.

\bibitem{passerat2020blockchain}
Jonathan Passerat-Palmbach, Tyler Farnan, Mike McCoy, Justin~D Harris, Sean~T
  Manion, Heather~Leigh Flannery, and Bill Gleim.
\newblock Blockchain-orchestrated machine learning for privacy preserving
  federated learning in electronic health data.
\newblock In {\em 2020 IEEE International Conference on Blockchain
  (Blockchain)}, pages 550--555. IEEE, 2020.

\bibitem{pentyala2022privfairfl}
Sikha Pentyala, Nicola Neophytou, Anderson Nascimento, Martine~De Cock, and
  Golnoosh Farnadi.
\newblock Privfairfl: Privacy-preserving group fairness in federated learning,
  2022.

\bibitem{putzel2022blackbox}
Preston Putzel and Scott Lee.
\newblock Blackbox post-processing for multiclass fairness.
\newblock {\em arXiv preprint arXiv:2201.04461}, 2022.

\bibitem{qi2022fairvfl}
Tao Qi, Fangzhao Wu, Chuhan Wu, Lingjuan Lyu, Tong Xu, Zhongliang Yang,
  Yongfeng Huang, and Xing Xie.
\newblock Fairvfl: A fair vertical federated learning framework with
  contrastive adversarial learning, 2022.

\bibitem{9291636}
Yang Qin, Hiroki Matsutani, and Masaaki Kondo.
\newblock A selective model aggregation approach in federated learning for
  online anomaly detection.
\newblock In {\em 2020 International Conferences on Internet of Things
  (iThings) and IEEE Green Computing and Communications (GreenCom) and IEEE
  Cyber, Physical and Social Computing (CPSCom) and IEEE Smart Data (SmartData)
  and IEEE Congress on Cybermatics (Cybermatics)}, pages 684--691, 2020.

\bibitem{qu2021contextaware}
Zhe Qu, Rui Duan, Lixing Chen, Jie Xu, Zhuo Lu, and Yao Liu.
\newblock Context-aware online client selection for hierarchical federated
  learning, 2021.

\bibitem{reddi2021adaptive}
Sashank Reddi, Zachary Charles, Manzil Zaheer, Zachary Garrett, Keith Rush,
  Jakub Konečný, Sanjiv Kumar, and H.~Brendan McMahan.
\newblock Adaptive federated optimization, 2021.

\bibitem{ribero2022federating}
M{\'o}nica Ribero, Jette Henderson, Sinead Williamson, and Haris Vikalo.
\newblock Federating recommendations using differentially private prototypes.
\newblock {\em Pattern Recognition}, 129:108746, 2022.

\bibitem{rodriguez2020federated}
Nuria Rodr{\'\i}guez-Barroso, Goran Stipcich, Daniel Jim{\'e}nez-L{\'o}pez,
  Jos{\'e}~Antonio Ruiz-Mill{\'a}n, Eugenio Mart{\'\i}nez-C{\'a}mara, Gerardo
  Gonz{\'a}lez-Seco, M~Victoria Luz{\'o}n, Miguel~Angel Veganzones, and
  Francisco Herrera.
\newblock Federated learning and differential privacy: Software tools analysis,
  the sherpa. ai fl framework and methodological guidelines for preserving data
  privacy.
\newblock {\em Information Fusion}, 64:270--292, 2020.

\bibitem{rodríguezgálvez2022enforcing}
Borja Rodríguez-Gálvez, Filip Granqvist, Rogier van Dalen, and Matt Seigel.
\newblock Enforcing fairness in private federated learning via the modified
  method of differential multipliers, 2022.

\bibitem{saha2021federated}
Sudipan Saha and Tahir Ahmad.
\newblock Federated transfer learning: concept and applications, 2021.

\bibitem{salazar2022fairfate}
Teresa Salazar, Miguel Fernandes, Helder Araujo, and Pedro~Henriques Abreu.
\newblock Fair-fate: Fair federated learning with momentum, 2022.

\bibitem{seif2020wireless}
Mohamed Seif, Ravi Tandon, and Ming Li.
\newblock Wireless federated learning with local differential privacy.
\newblock In {\em 2020 IEEE International Symposium on Information Theory
  (ISIT)}, pages 2604--2609. IEEE, 2020.

\bibitem{10.1145/359168.359176}
Adi Shamir.
\newblock How to share a secret.
\newblock {\em Commun. ACM}, 22(11):612–613, nov 1979.

\bibitem{sharma2022federated}
Pranay Sharma, Rohan Panda, Gauri Joshi, and Pramod~K. Varshney.
\newblock Federated minimax optimization: Improved convergence analyses and
  algorithms, 2022.

\bibitem{Shi_2023}
Yuxin Shi, Han Yu, and Cyril Leung.
\newblock Towards fairness-aware federated learning.
\newblock {\em {IEEE} Transactions on Neural Networks and Learning Systems},
  pages 1--17, 2023.

\bibitem{shokri2015privacy}
Reza Shokri and Vitaly Shmatikov.
\newblock Privacy-preserving deep learning.
\newblock In {\em Proceedings of the 22nd ACM SIGSAC conference on computer and
  communications security}, pages 1310--1321, 2015.

\bibitem{5958033}
Reza Shokri, George Theodorakopoulos, Jean-Yves Le~Boudec, and Jean-Pierre
  Hubaux.
\newblock Quantifying location privacy.
\newblock In {\em 2011 IEEE Symposium on Security and Privacy}, pages 247--262,
  2011.

\bibitem{shyn2021fedccea}
Sung~Kuk Shyn, Donghee Kim, and Kwangsu Kim.
\newblock Fedccea : A practical approach of client contribution evaluation for
  federated learning, 2021.

\bibitem{small2023equalised}
Edward~A Small, Kacper Sokol, Daniel Manning, Flora~D Salim, and Jeffrey Chan.
\newblock Equalised odds is not equal individual odds: Post-processing for
  group and individual fairness.
\newblock {\em arXiv preprint arXiv:2304.09779}, 2023.

\bibitem{9109557}
Mengkai Song, Zhibo Wang, Zhifei Zhang, Yang Song, Qian Wang, Ju Ren, and
  Hairong Qi.
\newblock Analyzing user-level privacy attack against federated learning.
\newblock {\em IEEE Journal on Selected Areas in Communications},
  38(10):2430--2444, 2020.

\bibitem{9006327}
Tianshu Song, Yongxin Tong, and Shuyue Wei.
\newblock Profit allocation for federated learning.
\newblock In {\em 2019 IEEE International Conference on Big Data (Big Data)},
  pages 2577--2586, 2019.

\bibitem{9458510}
Tianqi Su, Meiqi Wang, and Zhongfeng Wang.
\newblock Federated regularization learning: an accurate and safe method for
  federated learning.
\newblock In {\em 2021 IEEE 3rd International Conference on Artificial
  Intelligence Circuits and Systems (AICAS)}, pages 1--4, 2021.

\bibitem{sun2020ldp}
Lichao Sun, Jianwei Qian, and Xun Chen.
\newblock Ldp-fl: Practical private aggregation in federated learning with
  local differential privacy.
\newblock {\em arXiv preprint arXiv:2007.15789}, 2020.

\bibitem{sweeney2002k}
Latanya Sweeney.
\newblock k-anonymity: A model for protecting privacy.
\newblock {\em International journal of uncertainty, fuzziness and
  knowledge-based systems}, 10(05):557--570, 2002.

\bibitem{triastcyn2019federated}
Aleksei Triastcyn and Boi Faltings.
\newblock Federated learning with bayesian differential privacy.
\newblock In {\em 2019 IEEE International Conference on Big Data (Big Data)},
  pages 2587--2596. IEEE, 2019.

\bibitem{truex2019hybrid}
Stacey Truex, Nathalie Baracaldo, Ali Anwar, Thomas Steinke, Heiko Ludwig, Rui
  Zhang, and Yi Zhou.
\newblock A hybrid approach to privacy-preserving federated learning.
\newblock In {\em Proceedings of the 12th ACM workshop on artificial
  intelligence and security}, pages 1--11, 2019.

\bibitem{10.1145/3378679.3394533}
Stacey Truex, Ling Liu, Ka-Ho Chow, Mehmet~Emre Gursoy, and Wenqi Wei.
\newblock Ldp-fed: Federated learning with local differential privacy.
\newblock In {\em Proceedings of the Third ACM International Workshop on Edge
  Systems, Analytics and Networking}, EdgeSys '20, page 61–66, New York, NY,
  USA, 2020. Association for Computing Machinery.

\bibitem{tu2021incentive}
Xuezhen Tu, Kun Zhu, Nguyen~Cong Luong, Dusit Niyato, Yang Zhang, and Juan Li.
\newblock Incentive mechanisms for federated learning: From economic and game
  theoretic perspective, 2021.

\bibitem{usmanova2021distillation}
Anastasiia Usmanova, Fran{\c{c}}ois Portet, Philippe Lalanda, and German Vega.
\newblock A distillation-based approach integrating continual learning and
  federated learning for pervasive services.
\newblock {\em arXiv preprint arXiv:2109.04197}, 2021.

\bibitem{Wagner_2017}
Isabel Wagner.
\newblock Evaluating the strength of genomic privacy metrics.
\newblock {\em {ACM} Transactions on Privacy and Security}, 20(1):1--34, jan
  2017.

\bibitem{Wagner_2018}
Isabel Wagner and David Eckhoff.
\newblock Technical privacy metrics.
\newblock {\em {ACM} Computing Surveys}, 51(3):1--38, jun 2018.

\bibitem{wang2020hybrid}
Chang Wang, Jian Liang, Mingkai Huang, Bing Bai, Kun Bai, and Hao Li.
\newblock Hybrid differentially private federated learning on vertically
  partitioned data, 2020.

\bibitem{wang2022safeguarding}
Chen Wang, Xinkui Wu, Gaoyang Liu, Tianping Deng, Kai Peng, and Shaohua Wan.
\newblock Safeguarding cross-silo federated learning with local differential
  privacy.
\newblock {\em Digital Communications and Networks}, 8(4):446--454, 2022.

\bibitem{wang2019interpret}
Guan Wang.
\newblock Interpret federated learning with shapley values.
\newblock {\em arXiv preprint arXiv:1905.04519}, 2019.

\bibitem{9006179}
Guan Wang, Charlie~Xiaoqian Dang, and Ziye Zhou.
\newblock Measure contribution of participants in federated learning.
\newblock In {\em 2019 IEEE International Conference on Big Data (Big Data)},
  pages 2597--2604, 2019.

\bibitem{9155494}
Hao Wang, Zakhary Kaplan, Di Niu, and Baochun Li.
\newblock Optimizing federated learning on non-iid data with reinforcement
  learning.
\newblock In {\em IEEE INFOCOM 2020 - IEEE Conference on Computer
  Communications}, pages 1698--1707, 2020.

\bibitem{wang2020federated}
Hongyi Wang, Mikhail Yurochkin, Yuekai Sun, Dimitris Papailiopoulos, and
  Yasaman Khazaeni.
\newblock Federated learning with matched averaging, 2020.

\bibitem{wang2023fedeba}
Lin Wang, Zhichao Wang, and Xiaoying Tang.
\newblock Fedeba+: Towards fair and effective federated learning via
  entropy-based model, 2023.

\bibitem{wang2021federated}
Zheng Wang, Xiaoliang Fan, Jianzhong Qi, Chenglu Wen, Cheng Wang, and Rongshan
  Yu.
\newblock Federated learning with fair averaging, 2021.

\bibitem{wang2018inferring}
Zhibo Wang, Mengkai Song, Zhifei Zhang, Yang Song, Qian Wang, and Hairong Qi.
\newblock Beyond inferring class representatives: User-level privacy leakage
  from federated learning, 2018.

\bibitem{wang2019beyond}
Zhibo Wang, Mengkai Song, Zhifei Zhang, Yang Song, Qian Wang, and Hairong Qi.
\newblock Beyond inferring class representatives: User-level privacy leakage
  from federated learning.
\newblock In {\em IEEE INFOCOM 2019-IEEE conference on computer
  communications}, pages 2512--2520. IEEE, 2019.

\bibitem{wei2020federated}
Kang Wei, Jun Li, Ming Ding, Chuan Ma, Howard~H Yang, Farhad Farokhi, Shi Jin,
  Tony~QS Quek, and H~Vincent Poor.
\newblock Federated learning with differential privacy: Algorithms and
  performance analysis.
\newblock {\em IEEE Transactions on Information Forensics and Security},
  15:3454--3469, 2020.

\bibitem{wei2022vertical}
Kang Wei, Jun Li, Chuan Ma, Ming Ding, Sha Wei, Fan Wu, Guihai Chen, and
  Thilina Ranbaduge.
\newblock Vertical federated learning: Challenges, methodologies and
  experiments.
\newblock {\em arXiv preprint arXiv:2202.04309}, 2022.

\bibitem{wu2020towards}
Danye Wu, Miao Pan, Zhiwei Xu, Yujun Zhang, and Zhu Han.
\newblock Towards efficient secure aggregation for model update in federated
  learning.
\newblock In {\em GLOBECOM 2020-2020 IEEE Global Communications Conference},
  pages 1--6. IEEE, 2020.

\bibitem{wu2022defense}
Jing Wu, Munawar Hayat, Mingyi Zhou, and Mehrtash Harandi.
\newblock Defense against privacy leakage in federated learning, 2022.

\bibitem{wu2022motley}
Shanshan Wu, Tian Li, Zachary Charles, Yu Xiao, Ziyu Liu, Zheng Xu, and
  Virginia Smith.
\newblock Motley: Benchmarking heterogeneity and personalization in federated
  learning, 2022.

\bibitem{wu2022adaptive}
Xiang Wu, Yongting Zhang, Minyu Shi, Pei Li, Ruirui Li, and Neal~N Xiong.
\newblock An adaptive federated learning scheme with differential privacy
  preserving.
\newblock {\em Future Generation Computer Systems}, 127:362--372, 2022.

\bibitem{xi2021batfl}
Binhan Xi, Shaofeng Li, Jiachun Li, Hui Liu, Hong Liu, and Haojin Zhu.
\newblock Batfl: Backdoor detection on federated learning in e-health.
\newblock In {\em 2021 IEEE/ACM 29th International Symposium on Quality of
  Service (IWQOS)}, pages 1--10. IEEE, 2021.

\bibitem{xia2021vertical}
Wensheng Xia, Ying Li, Lan Zhang, Zhonghai Wu, and Xiaoyong Yuan.
\newblock A vertical federated learning framework for horizontally partitioned
  labels, 2021.

\bibitem{9142401}
Wenchao Xia, Tony Q.~S. Quek, Kun Guo, Wanli Wen, Howard~H. Yang, and Hongbo
  Zhu.
\newblock Multi-armed bandit-based client scheduling for federated learning.
\newblock {\em IEEE Transactions on Wireless Communications},
  19(11):7108--7123, 2020.

\bibitem{xie2021efficient}
Lunchen Xie, Jiaqi Liu, Songtao Lu, Tsung hui Chang, and Qingjiang Shi.
\newblock An efficient learning framework for federated xgboost using secret
  sharing and distributed optimization, 2021.

\bibitem{xin2020private}
Bangzhou Xin, Wei Yang, Yangyang Geng, Sheng Chen, Shaowei Wang, and Liusheng
  Huang.
\newblock Private fl-gan: Differential privacy synthetic data generation based
  on federated learning.
\newblock In {\em ICASSP 2020-2020 IEEE International Conference on Acoustics,
  Speech and Signal Processing (ICASSP)}, pages 2927--2931. IEEE, 2020.

\bibitem{8765347}
Guowen Xu, Hongwei Li, Sen Liu, Kan Yang, and Xiaodong Lin.
\newblock Verifynet: Secure and verifiable federated learning.
\newblock {\em IEEE Transactions on Information Forensics and Security},
  15:911--926, 2020.

\bibitem{xu2019hybridalpha}
Runhua Xu, Nathalie Baracaldo, Yi Zhou, Ali Anwar, and Heiko Ludwig.
\newblock Hybridalpha: An efficient approach for privacy-preserving federated
  learning.
\newblock In {\em Proceedings of the 12th ACM workshop on artificial
  intelligence and security}, pages 13--23, 2019.

\bibitem{xu2021reputation}
Xinyi Xu and Lingjuan Lyu.
\newblock A reputation mechanism is all you need: Collaborative fairness and
  adversarial robustness in federated learning, 2021.

\bibitem{yang2020fairness}
Ke Yang, Biao Huang, Julia Stoyanovich, and Sebastian Schelter.
\newblock Fairness-aware instrumentation of preprocessing\~{} pipelines for
  machine learning.
\newblock In {\em Workshop on Human-In-the-Loop Data Analytics (HILDA'20)},
  2020.

\bibitem{yang2019parallel}
Shengwen Yang, Bing Ren, Xuhui Zhou, and Liping Liu.
\newblock Parallel distributed logistic regression for vertical federated
  learning without third-party coordinator, 2019.

\bibitem{yang2021accuracylossless}
Xue Yang, Yan Feng, Weijun Fang, Jun Shao, Xiaohu Tang, Shu-Tao Xia, and
  Rongxing Lu.
\newblock An accuracy-lossless perturbation method for defending privacy
  attacks in federated learning, 2021.

\bibitem{yang2021learning}
Xiulong Yang and Shihao Ji.
\newblock Learning with multiplicative perturbations.
\newblock In {\em 2020 25th International Conference on Pattern Recognition
  (ICPR)}, pages 1321--1328. IEEE, 2021.

\bibitem{9861177}
Yurui Yang and Bo Jiang.
\newblock Towards group fairness via semi-centralized adversarial training in
  federated learning.
\newblock In {\em 2022 23rd IEEE International Conference on Mobile Data
  Management (MDM)}, pages 482--487, 2022.

\bibitem{yang2020fpgabased}
Zhaoxiong Yang, Shuihai Hu, and Kai Chen.
\newblock Fpga-based hardware accelerator of homomorphic encryption for
  efficient federated learning, 2020.

\bibitem{9260194}
Yunfan Ye, Shen Li, Fang Liu, Yonghao Tang, and Wanting Hu.
\newblock Edgefed: Optimized federated learning based on edge computing.
\newblock {\em IEEE Access}, 8:209191--209198, 2020.

\bibitem{10.1145/3460427}
Xuefei Yin, Yanming Zhu, and Jiankun Hu.
\newblock A comprehensive survey of privacy-preserving federated learning: A
  taxonomy, review, and future directions.
\newblock {\em ACM Comput. Surv.}, 54(6), jul 2021.

\bibitem{10.1145/3375627.3375840}
Han Yu, Zelei Liu, Yang Liu, Tianjian Chen, Mingshu Cong, Xi Weng, Dusit
  Niyato, and Qiang Yang.
\newblock A fairness-aware incentive scheme for federated learning.
\newblock In {\em Proceedings of the AAAI/ACM Conference on AI, Ethics, and
  Society}, AIES '20, page 393–399, New York, NY, USA, 2020. Association for
  Computing Machinery.

\bibitem{yu2020fairness}
Han Yu, Zelei Liu, Yang Liu, Tianjian Chen, Mingshu Cong, Xi Weng, Dusit
  Niyato, and Qiang Yang.
\newblock A fairness-aware incentive scheme for federated learning.
\newblock In {\em Proceedings of the AAAI/ACM Conference on AI, Ethics, and
  Society}, pages 393--399, 2020.

\bibitem{yue2022gifairfl}
Xubo Yue, Maher Nouiehed, and Raed~Al Kontar.
\newblock Gifair-fl: A framework for group and individual fairness in federated
  learning, 2022.

\bibitem{10004844}
Won~Joon Yun, Yunseok Kwak, Hankyul Baek, Soyi Jung, Mingyue Ji, Mehdi Bennis,
  Jihong Park, and Joongheon Kim.
\newblock Slimfl: Federated learning with superposition coding over slimmable
  neural networks.
\newblock {\em IEEE/ACM Transactions on Networking}, pages 1--16, 2022.

\bibitem{zeng2022improving}
Yuchen Zeng, Hongxu Chen, and Kangwook Lee.
\newblock Improving fairness via federated learning, 2022.

\bibitem{zhang2018mitigating}
Brian~Hu Zhang, Blake Lemoine, and Margaret Mitchell.
\newblock Mitigating unwanted biases with adversarial learning.
\newblock In {\em Proceedings of the 2018 AAAI/ACM Conference on AI, Ethics,
  and Society}, pages 335--340, 2018.

\bibitem{zhang2020batchcrypt}
Chengliang Zhang, Suyi Li, Junzhe Xia, Wei Wang, Feng Yan, and Yang Liu.
\newblock Batchcrypt: Efficient homomorphic encryption for cross-silo federated
  learning.
\newblock In {\em Proceedings of the 2020 USENIX Annual Technical Conference
  (USENIX ATC 2020)}, 2020.

\bibitem{9378043}
Daniel~Yue Zhang, Ziyi Kou, and Dong Wang.
\newblock Fairfl: A fair federated learning approach to reducing demographic
  bias in privacy-sensitive classification models.
\newblock In {\em 2020 IEEE International Conference on Big Data (Big Data)},
  pages 1051--1060, 2020.

\bibitem{zhang2022unified}
Fengda Zhang, Kun Kuang, Yuxuan Liu, Long Chen, Chao Wu, Fei Wu, Jiaxun Lu,
  Yunfeng Shao, and Jun Xiao.
\newblock Unified group fairness on federated learning, 2022.

\bibitem{zhang2021omnifair}
Hantian Zhang, Xu Chu, Abolfazl Asudeh, and Shamkant~B Navathe.
\newblock Omnifair: A declarative system for model-agnostic group fairness in
  machine learning.
\newblock In {\em Proceedings of the 2021 international conference on
  management of data}, pages 2076--2088, 2021.

\bibitem{zhang2023privacy}
Haobo Zhang, Junyuan Hong, Fan Dong, Steve Drew, Liangjie Xue, and Jiayu Zhou.
\newblock A privacy-preserving hybrid federated learning framework for
  financial crime detection.
\newblock {\em arXiv preprint arXiv:2302.03654}, 2023.

\bibitem{zhang2019pefl}
Jiale Zhang, Bing Chen, Shui Yu, and Hai Deng.
\newblock Pefl: A privacy-enhanced federated learning scheme for big data
  analytics.
\newblock In {\em 2019 IEEE Global Communications Conference (GLOBECOM)}, pages
  1--6. IEEE, 2019.

\bibitem{zhang2022dense}
Jie Zhang, Chen Chen, Bo Li, Lingjuan Lyu, Shuang Wu, Shouhong Ding, Chunhua
  Shen, and Chao Wu.
\newblock Dense: Data-free one-shot federated learning.
\newblock {\em Advances in Neural Information Processing Systems},
  35:21414--21428, 2022.

\bibitem{zhang2021practical}
Jie Zhang, Chen Chen, Bo Li, Lingjuan Lyu, Shuang Wu, Jianghe Xu, Shouhong
  Ding, and Chao Wu.
\newblock A practical data-free approach to one-shot federated learning with
  heterogeneity.
\newblock {\em arXiv preprint arXiv:2112.12371}, 2021.

\bibitem{zhang2022privacy}
Jingyang Zhang, Yiran Chen, and Hai Li.
\newblock Privacy leakage of adversarial training models in federated learning
  systems, 2022.

\bibitem{286415}
Junxue Zhang, Xiaodian Cheng, Wei Wang, Liu Yang, Jinbin Hu, and Kai Chen.
\newblock {FLASH}: Towards a high-performance hardware acceleration
  architecture for cross-silo federated learning.
\newblock In {\em 20th USENIX Symposium on Networked Systems Design and
  Implementation (NSDI 23)}, pages 1057--1079, Boston, MA, Apr. 2023. USENIX
  Association.

\bibitem{zhang2020hierarchically}
Jingfeng Zhang, Cheng Li, Antonio Robles-Kelly, and Mohan Kankanhalli.
\newblock Hierarchically fair federated learning, 2020.

\bibitem{9812492}
Li Zhang, Jianbo Xu, Pandi Vijayakumar, Pradip~Kumar Sharma, and Uttam Ghosh.
\newblock Homomorphic encryption-based privacy-preserving federated learning in
  iot-enabled healthcare system.
\newblock {\em IEEE Transactions on Network Science and Engineering}, pages
  1--17, 2022.

\bibitem{10.1145/3472456.3473513}
Shulai Zhang, Zirui Li, Quan Chen, Wenli Zheng, Jingwen Leng, and Minyi Guo.
\newblock Dubhe: Towards data unbiasedness with homomorphic encryption in
  federated learning client selection.
\newblock In {\em Proceedings of the 50th International Conference on Parallel
  Processing}, ICPP '21, New York, NY, USA, 2021. Association for Computing
  Machinery.

\bibitem{zhang2022multiagent}
Sai~Qian Zhang, Jieyu Lin, and Qi Zhang.
\newblock A multi-agent reinforcement learning approach for efficient client
  selection in federated learning, 2022.

\bibitem{zhang2022understanding}
Xinwei Zhang, Xiangyi Chen, Mingyi Hong, Zhiwei~Steven Wu, and Jinfeng Yi.
\newblock Understanding clipping for federated learning: Convergence and
  client-level differential privacy.
\newblock In {\em International Conference on Machine Learning, ICML 2022},
  2022.

\bibitem{9908546}
Xiao-Yu Zhang, José-Rodrigo Córdoba-Pachón, Peiqian Guo, Chris Watkins, and
  Stefanie Kuenzel.
\newblock Privacy-preserving federated learning for value-added service model
  in advanced metering infrastructure.
\newblock {\em IEEE Transactions on Computational Social Systems}, pages 1--15,
  2022.

\bibitem{10.1145/3457388.3458665}
Yuhui Zhang, Zhiwei Wang, Jiangfeng Cao, Rui Hou, and Dan Meng.
\newblock Shufflefl: Gradient-preserving federated learning using trusted
  execution environment.
\newblock In {\em Proceedings of the 18th ACM International Conference on
  Computing Frontiers}, CF '21, page 161–168, New York, NY, USA, 2021.
  Association for Computing Machinery.

\bibitem{9325934}
Bin Zhao, Kai Fan, Kan Yang, Zilong Wang, Hui Li, and Yintang Yang.
\newblock Anonymous and privacy-preserving federated learning with industrial
  big data.
\newblock {\em IEEE Transactions on Industrial Informatics}, 17(9):6314--6323,
  2021.

\bibitem{zhao2023truthful}
Yuxi Zhao, Xiaowen Gong, and Shiwen Mao.
\newblock Truthful incentive mechanism for federated learning with crowdsourced
  data labeling, 2023.

\bibitem{zhao2020local}
Yang Zhao, Jun Zhao, Mengmeng Yang, Teng Wang, Ning Wang, Lingjuan Lyu, Dusit
  Niyato, and Kwok-Yan Lam.
\newblock Local differential privacy-based federated learning for internet of
  things.
\newblock {\em IEEE Internet of Things Journal}, 8(11):8836--8853, 2020.

\bibitem{zheng2022secure}
Shuyuan Zheng, Yang Cao, and Masatoshi Yoshikawa.
\newblock Secure shapley value for cross-silo federated learning.
\newblock {\em arXiv preprint arXiv:2209.04856}, 2022.

\bibitem{10.1145/3534585}
Pengyuan Zhou, Hengwei Xu, Lik~Hang Lee, Pei Fang, and Pan Hui.
\newblock Are you left out? an efficient and fair federated learning for
  personalized profiles on wearable devices of inferior networking conditions.
\newblock {\em Proc. ACM Interact. Mob. Wearable Ubiquitous Technol.}, 6(2),
  jul 2022.

\bibitem{9814669}
Hongbin Zhu, Miao Yang, Junqian Kuang, Hua Qian, and Yong Zhou.
\newblock Client selection for asynchronous federated learning with fairness
  consideration.
\newblock In {\em 2022 IEEE International Conference on Communications
  Workshops (ICC Workshops)}, pages 800--805, 2022.

\bibitem{9916164}
Hongbin Zhu, Yong Zhou, Hua Qian, Yuanming Shi, Xu Chen, and Yang Yang.
\newblock Online client selection for asynchronous federated learning with
  fairness consideration.
\newblock {\em IEEE Transactions on Wireless Communications}, 22(4):2493--2506,
  2023.

\bibitem{zhu2019deep}
Ligeng Zhu, Zhijian Liu, and Song Han.
\newblock Deep leakage from gradients, 2019.

\bibitem{zhu2021data}
Zhuangdi Zhu, Junyuan Hong, and Jiayu Zhou.
\newblock Data-free knowledge distillation for heterogeneous federated
  learning.
\newblock In {\em International Conference on Machine Learning}, pages
  12878--12889. PMLR, 2021.

\end{thebibliography}
}

\end{document}